\documentclass[12pt]{article}

\newif\ifdraft
\draftfalse

\ifdraft
  \usepackage[left=0.1in,right = 1.9in,textwidth=6.5in,top=1in,bottom=1in,marginparwidth=1.4in,marginparsep=0.15in]{geometry}
\else
  \usepackage[margin=1in]{geometry}
\fi
\usepackage{graphicx}
\usepackage{setspace}
\usepackage{arydshln}
\usepackage[all]{xy}
\usepackage{amsmath}
\usepackage{amssymb}
\usepackage{amsthm}
\DeclareFontFamily{U}{stmry}{}
\DeclareFontShape{U}{stmry}{m}{n}{<-> stmary10}{}
\DeclareSymbolFont{stmry}{U}{stmry}{m}{n}
\DeclareMathSymbol{\shortrightarrow}{\mathrel}{stmry}{"01}

\usepackage{bm}
\usepackage{booktabs}
\usepackage{array}
\usepackage{float}
\usepackage{longtable}
\usepackage{calc}
\usepackage{tabularx} 
\usepackage{array}    
\usepackage{adjustbox}
\usepackage[normalem]{ulem}
\usepackage{enumitem}
\usepackage{tikz}
\usepackage{accents}
\usepackage{xcolor}
\usepackage{enumitem}

\usepackage{caption}

\colorlet{rsiHuman}{red!80!black}       
\colorlet{rsiAlgo}{blue!70!black}       
\colorlet{rsiTrain}{orange!85!black}    
\colorlet{rsiCap}{purple!80!black}      
\colorlet{rsiEcon}{green!50!black}      

\newcommand{\adot}[1]{\accentset{\scalebox{0.3}{$\bullet$}}{#1}}

\usepackage[most]{tcolorbox}
\usepackage{amsmath}
\usepackage{amssymb}

\usepackage[authoryear,round]{natbib}
\newcommand{\paperbibliography}{%
  {\footnotesize
  \setlength{\bibsep}{0pt}
  \bibliographystyle{plainnat}
  \bibliography{references}}%
}

\usepackage{array}
\usepackage{booktabs}
\usepackage{colortbl}
\usepackage{graphicx}
\usepackage{longtable}
\usepackage{tabularx}
\usepackage{wrapfig2} 
\usepackage{needspace} 
\newcommand{\rsiscale}{0.72}
\usepackage{caption}
\usepackage{capt-of}
\usepackage{tikz}
\usetikzlibrary{calc,positioning,arrows.meta,matrix}

\newcolumntype{Y}{>{\vspace{0pt}\raggedright\arraybackslash}X}

\tikzset{
    rsi diagram/.style={
        node distance=0.5cm and 1.5cm,
        every path/.style={->},
        every node/.style={rsi node}
    },
    rsi node/.style={draw, rectangle, inner sep=4pt, outer sep=2pt},
    rd/.style={blue!70!black, very thick},
    humanlabor/.style={red!80!black, very thick}, 
    training/.style={orange!85!black, very thick},
    uplift/.style={purple!80!black, very thick},
    economic/.style={green!50!black, very thick},
    inferencecompute/.style={teal!80!black, very thick},
    experimentcompute/.style={brown!80!black, very thick},
    datacol/.style={magenta!70!black, very thick},
    rsi edge label/.style={draw=none, fill=white, inner sep=1pt, font=\small},
    rate node/.style={draw, rounded corners=3pt, inner sep=4pt, align=center},
    cg edge/.style={->, rounded corners=4pt},
    cg trunk/.style={-, rounded corners=4pt}
}

\newcommand{\rsiNode}[3][]{\node[#1] (#2) {#3};}

\definecolor{newblue}{rgb}{0.16796875,0.396484375,0.669921875}
\usepackage[unicode=true,  bookmarks=true, bookmarksnumbered=false, bookmarksopen=false, breaklinks=true, pdfborder={0 0 0}, pdfborderstyle={}, backref=false, colorlinks=true]{hyperref}
\usepackage{xurl}
\hypersetup{
    colorlinks=true,       
    linkcolor=newblue,          
    citecolor=black,        
    urlcolor=black           
}

\usetikzlibrary{calc,intersections,matrix,positioning, shapes, shapes.geometric, arrows, arrows.meta}

\usepackage{subfiles} 
\usepackage{tablefootnote} 
\newcolumntype{C}{>{\centering\arraybackslash}X}
\usepackage{mathpazo} 
\usepackage{titlesec}
\usepackage{marginnote}
\usepackage{epigraph}
\ifdraft
  \newcommand*{\draftmarginfont}{\normalfont\normalsize\color{black}}
  \newcommand{\draftmarginbox}[1]{%
    \colorbox{yellow}{%
      \parbox[t]{\dimexpr\marginparwidth-2\fboxsep\relax}{%
        \raggedright\draftmarginfont #1%
      }%
    }%
  }
  
  \let\plainmarginnote\marginnote
  \RenewDocumentCommand{\marginnote}{o +m o}{%
    \IfNoValueTF{#3}{%
      \IfNoValueTF{#1}{%
        \plainmarginnote{\draftmarginbox{#2}}%
      }{%
        \plainmarginnote[\draftmarginbox{#1}]{\draftmarginbox{#2}}%
      }%
    }{%
      \IfNoValueTF{#1}{%
        \plainmarginnote{\draftmarginbox{#2}}[#3]%
      }{%
        \plainmarginnote[\draftmarginbox{#1}]{\draftmarginbox{#2}}[#3]%
      }%
    }%
  }
  \let\plainmarginpar\marginpar
  \RenewDocumentCommand{\marginpar}{o +m}{%
    \IfNoValueTF{#1}{%
      \plainmarginpar{\draftmarginbox{#2}}%
    }{%
      \plainmarginpar[\draftmarginbox{#1}]{\draftmarginbox{#2}}%
    }%
  }
\else
  \renewcommand{\marginnote}[2][]{}
  \RenewDocumentCommand{\marginpar}{o +m}{}
\fi

\titleformat{\section}
		{\Large\bfseries\center}
         {\thesection}
        {0.5em}
        {\textsc}
        []

\titleformat{\subsection}
        {\large\bfseries}
        {\thesubsection}
        {0.5em}
\allowdisplaybreaks

\theoremstyle{plain}

\theoremstyle{definition}

\theoremstyle{remark}

\usepackage{tikz}
\usetikzlibrary{calc,intersections,matrix,positioning, shapes, shapes.geometric, arrows, arrows.meta}
\tikzset{ebox/.style={draw,rectangle split,rectangle split horizontal,rectangle split parts=2,text centered}}
\tikzset{
    rsi diagram/.style={
        node distance=0.5cm and 1.5cm,
        every path/.style={->},
        every node/.style={rsi node}
    },
    rsi node/.style={draw, rectangle, inner sep=4pt, outer sep=2pt},
    rd/.style={blue!70!black, very thick},
    humanlabor/.style={red!80!black, very thick},
    training/.style={orange!85!black, very thick},
    uplift/.style={purple!80!black, very thick},
    rsi edge label/.style={draw=none, fill=white, inner sep=1pt, font=\small},
    legend label/.style={draw=none, inner sep=1pt},
    economic/.style={green!50!black, very thick},
}
\providecommand{\rsiNode}[3][]{\node[#1] (#2) {#3};}

\title{The Economics of Recursive Self-Improvement}
\author{%
Tom Cunningham\textsuperscript{*} \\[0.6em]
Lukas Althoff\textsuperscript{\dag} \quad 
Basil Halperin\textsuperscript{\dag} \quad
Brian Jabarian\textsuperscript{\dag} \quad
Andrew Koh\textsuperscript{\dag} \\[0.3em]
Arjun Ramani\textsuperscript{\dag} \quad
Phil Trammell\textsuperscript{\dag} \quad 
Parker Whitfill\textsuperscript{\dag} \quad
Cheryl Wu\textsuperscript{\dag}%
}
\date{\today}

\begin{document}

\maketitle
\begingroup
\renewcommand{\thefootnote}{}%
\makeatletter
\renewcommand{\@makefntext}[1]{\noindent #1}%
\makeatother
\footnotetext{\textbf{Contributions:} \textsuperscript{*}Lead author.\quad \textsuperscript{\dag}Listed in alphabetical order; these authors contributed equally.\\
\textbf{Corresponding author:}  Tom Cunningham, \href{mailto:tom.cunningham@metr.org}{tom.cunningham@metr.org}\\
\textbf{Latest version of the paper: \url{https://github.com/elasticity-ai/elasticity/raw/main/paper/elasticity-rsi-paper.pdf}}\\
\textbf{Acknowledgements:} Thanks for comments from Anson Ho, Thomas Houlden, Shrey Jain, Whitney Zhang, Alan Chan, Thomas Kwa, Anton Korinek, Jason Abaluck.\\
\textbf{Affiliations}: Tom Cunningham (METR); Lukas Althoff (Stanford University); Basil Halperin (University of Virginia); Brian Jabarian (Carnegie Mellon University); Andrew Koh (Columbia University); Arjun Ramani (MIT); Phil Trammell (Stanford DEL and Epoch AI); Parker Whitfill (METR); Cheryl Wu (Yale University).
\\
\textit{All authors are affiliated with the Elasticity Institute (}\url{https://elasticity.institute}\textit{).}
\\
\textbf{Support:} We gratefully acknowledge administrative and financial support from METR, and hosting from Constellation.}%
\renewcommand{\thefootnote}{\arabic{footnote}}%
\endgroup

\begin{abstract}
\normalsize

We model the economics of recursive self-improvement (RSI) and assess its plausibility and impacts. 
First, we build a sequence of increasingly rich models of AI progress to highlight the feedback loops behind RSI.
We represent our models as directed graphs and show that net acceleration in AI capabilities depends on the product of elasticities across each feedback loop. Second, we distinguish between ``narrow'' and ``broad'' AI capabilities, capturing the possibility that AI systems improve narrowly at optimizing AI R\&D benchmarks without improving at broader economically valuable tasks. Third, we document existing estimates of key parameters, and provide a wish list of empirical objects that AI companies can measure and feasibly share publicly. Finally, we calibrate the model with existing data. A back-of-the-envelope calculation suggests that feedback loops are not currently strong enough to generate a self-sustaining acceleration, though they appear to be strengthening. We conclude by assessing the plausibility and implications of such an acceleration. 

\end{abstract}

\clearpage

\section{Introduction}

\paragraph{Motivation.} Over the past year, signs have emerged of a feedback loop in which AI systems speed up AI research itself, potentially accelerating the already rapid growth of AI capabilities \citep{favaro2026rsi}.\footnote{Measured, for example, via the Epoch Capabilities Index or the METR time horizon metric, though measuring capabilities remains difficult.} If this process is as strong as some expect, the resulting transformation could have consequences comparable in scale to historical shifts such as the Enlightenment, reshaping economic, social, and political life \citep{mokyr2002gifts}.


\paragraph{Core argument.} The degree of acceleration depends on what we call the \textit{core feedback loop}: for a one-unit increase in model capabilities, how much do the capabilities of the next generation of models increase? Estimating the strength of this relationship is challenging because it is governed by many inputs and possible bottlenecks. 

This note presents a series of theoretical models to clarify the forces behind the core feedback loop. We draw a tight connection to empirical data needed to measure the strength of the feedback loop, which can help assess the degree of current and future acceleration.


\paragraph{Defining RSI.}
The term ``recursive self-improvement'' (RSI) has been used in several ways, often inconsistently.\footnote{
The origin of the term is not clear, but it was popularized by (and plausibly originated with) \citet{yudkowsky2001creating, yudkowsky2008recursive}, extending an argument by \citet{good1965}.} Some define RSI as a technology contributing to its own improvement, which could apply to almost any technology since the dawn of humanity. \citet{favaro2026rsi} recently defined the term more narrowly as a technology that has fully automated the process of its own improvement. 
Other definitions highlight the importance of acceleration or feedback loops, even if full automation is not realized. 

\marginpar{\tiny TEC: I don't think ``agnostic'' is quite the right word. BJ: 'open to'? BH: `agnostic to the precise definition of the term RSI'? AR: `reworded a bit. feel free to update but i think good to be clear `why' we don't just call RSI a self-sustaining acceleration. BH: revised further} 

To avoid confusion, we instead focus on the possibility of a \textit{self-sustaining acceleration} in AI capabilities and derive the conditions under which it arises. We begin by defining self-sustaining acceleration and related terms in Table \ref{tab:definitions-references}.

\marginpar{\tiny AR: This feels like a pretty weak notion of automatability. Why ``unaided human researchers" and ``complementarities"? An optimizing firm then would then still employ humans... Why not ``A firm that is maximizing the rate of AI progress would choose not to use human AI R\&D researchers at any cost. BH: agreed, made two edits based on this but not your language, feel free to update}

\renewcommand{\arraystretch}{1.2}
\begin{longtable}{p{0.32\linewidth} p{0.64\linewidth}}

\caption{Definitions
}
\label{tab:definitions-references} \\
\toprule
\textbf{Term} & \textbf{Definition} \\
\midrule
\endfirsthead

\caption*{Table \thetable: Definitions} \\
\toprule
\textbf{Term} & \textbf{Definition} \\
\midrule
\endhead

\midrule
\endfoot

\bottomrule
\endlastfoot

Feedback loops &
 When outputs of a system today are routed back as inputs to the system tomorrow.
 Arguably most technologies exhibit feedback effects (or are part of feedback loops when embedded in the economy), and AI has certainly exhibited feedback loops for a long time.
 \\\addlinespace[0.9em]
 
R\&D automatability &
 When AI is able to autonomously make technological improvements at least equivalent to those made by human researchers, at an equal cost. Automatability does not imply automation: humans could still be employed in R\&D due to slow diffusion or regulation. 
 \\
\multicolumn{2}{@{}p{\linewidth}@{}}{%
\begin{tcolorbox}[
  colback=newblue!5,
  colframe=newblue!100,
  boxrule=0.6pt,
  arc=1mm,
  left=4pt,
  right=4pt,
  top=4pt,
  bottom=4pt,
  before skip=0pt,
  after skip=0pt
]
\begin{tabular}{@{}p{0.32\linewidth} p{0.64\linewidth}@{}}
Self-sustaining acceleration &
When AI systems are sufficient for accelerating progress in AI capabilities without any growth in exogenous inputs (human labor, training compute, etc.).
\end{tabular}
\end{tcolorbox}%
}
\\\addlinespace[0.9em]
 Intelligence explosion &
    When AI capabilities go to infinity in finite time.  
\\
\end{longtable}

\vspace{-1.5em} 
\noindent
\begin{minipage}{\textwidth}
	{\footnotesize\singlespacing \emph{Relevant references:} \citet{davidson2026automating}, \citet{chan2026measuring}, \citet{eth2025software}, \citet{davidsonhoulden2025}, \citet{davidson2023takeoff},  \citet{good1965}, \citet{hanson2008economics}, \citet{christiano2018takeoff}, \citet{ho2025experiments}, \citet{aghion2017artificial}.\par}
\end{minipage}
\vspace{1.5em}

These concepts are often conflated with each other and the term ``RSI.'' Some have treated full automation of AI R\&D as sufficient for self-sustaining acceleration. For example, I.J. Good in 1965 states: ``an ultraintelligent machine could design even better machines; there would then unquestionably be an intelligence explosion.'' But our models make it clear that such an explosion may not follow if there are diminishing returns (``ideas become harder to find'') or if feedback loops become bottlenecked.

\paragraph{Modeling RSI.} We construct a sequence of progressively richer models of RSI. We pre\-sent the models in diagrams for readability. 

Our basic model distinguishes two production functions that combine to form the core feedback loop: improvements to algorithmic efficiency are produced with human labor and AI capabilities; in turn, algorithmic efficiency, along with training compute, raises AI capabilities. A third production function, for economic output, determines how AI capabilities impact the wider economy. We extend the model to emphasize the possibility of bottlenecks, where the feedback loop may be broken by the necessity of humans, compute, or data; and for economic feedback loops, where higher output finances further compute investment and data collection, potentially alleviating bottlenecks. We derive a condition under which each model features a self-sustaining acceleration.

We also discuss the possibility that there would be an acceleration only in ``narrow'' capabilities. The core feedback loop requires a strong connection between algorithmic efficiency and the capabilities required to find new algorithmic optimizations. This connection could be strong without necessarily accelerating real-world impacts because such impacts depend on ``broad'' capabilities less sensitive to algorithmic efficiency. Finally, we discuss the possibility that AI capabilities may rapidly advance for specific optimization abilities, but not for all types of algorithmic improvements, bottlenecking RSI.

Our key technical contribution is to introduce a simple graphical framework to represent an otherwise complicated system of variables and feedback loops. Nodes of the graph represent outputs of a production function; the strengths of edges in the graph represent elasticities of an output with respect to inputs. There is a self-sustaining acceleration under a condition which can be derived simply using these elasticities and the graph. Details of this framework are outlined in technical boxes.

\paragraph{Existing empirical evidence \& requests for data.} Our models depend on a set of critical measurable parameters. We review existing empirical estimates of these parameters. 

We also put forward a list of measures that would be useful for AI companies to provide while remaining feasible to share publicly. The biggest unknowns to be informed by additional data are (i) the growth rate of algorithmic efficiency in labs, (ii) the fraction of lab R\&D expenditures going to each input (humans, data, experimental compute, inference for R\&D), and (iii) direct measures of how much models contribute to research, like the share of technical advances produced by AI. 

\paragraph{Plausibility and implications.} We make a tentative calibration of the self-sustaining acceleration condition using the existing data that is available, measuring AI capabilities using the Epoch Capabilities Index \citep{ho2025rosetta}. We find that the condition is met if a one-unit increase in AI model capabilities results in at least 15\% higher AI R\&D productivity. A rough back-of-the-envelope calculation based on reported AI engineer uplift suggests this return has been around 9\% since the launch of coding agents. This number is below the model-implied threshold, suggesting we are not experiencing a self-sustaining acceleration. Nonetheless, there is ample evidence that this return is not constant and has been increasing of late. Our model therefore does not rule out the possibility of self-sustaining acceleration in the near future.

Given the vast uncertainty in both the data and model behind our calibration, we also discuss other qualitative evidence for and against a self-sustaining acceleration. Existing survey and benchmark evidence suggests that AI systems are increasingly useful in discovering new algorithmic improvements. On the other hand, algorithmic progress has historically depended on continued compute scaling, and future compute growth may become constrained by power, capital, or broader economic growth. Moreover, even if the technical conditions identified by our calibration are eventually met, deployment could be substantially slowed by political and regulatory constraints. 

\paragraph{Position in the literature.} The economics literature on RSI is founded on \citet{aghion2017artificial}, who derived conditions under which AI-driven automation could generate accelerating or even explosive economic growth, and \citet{davidson2023takeoff}, whose com\-pute-centric framework offered an early formal model of AI takeoff dynamics. We complement an emerging literature on the economics of RSI in three ways. First, our networked models build on \citet{davidson2026automating}, who formally develop a general theory of semi-endogenous growth models with innovation networks plus economic feedback loops and apply it to AI. Second, 
we emphasize the distinction between narrow and broad capabilities in both theory and the discussion of economic impact. This distinction is largely set aside in the review articles of \citet{trammell2025economic} and \citet{jones2026ai}. Third, we provide a wish list of empirical objects that AI companies can both measure and feasibly share openly that would be informative for the public conversation on RSI. This supplements the existing empirical efforts that we review (e.g., \citealt{whitfill2025compute}, \citealt{gundlach2025origin}, \citealt{epoch2026key}).

\ifSubfilesClassLoaded{%
    \paperbibliography%
}{}

\end{document}

\section{Models of Recursive Self-Improvement}\label{sec:model}

\begingroup
\setlength{\abovedisplayskip}{4pt}
\setlength{\belowdisplayskip}{4pt}
\setlength{\parskip}{0.6em plus 0.15em}
\setlength{\parindent}{0pt}

\begin{sloppypar}
We present a progression of models to understand the economics of recursive self-improvement. The models are presented graphically, where each directed edge indicates that one variable is an input to the production function of the other, and with a strength that is an elasticity. The condition for self-sustaining acceleration can be read off from the graph: as we describe below, it is related to the partial elasticities represented by the strength of the edges. In technical details boxes, for interested readers only, we make precise how to translate our graphs into equations. 
\end{sloppypar}

\needspace{2.2in}
\subsection{Basic Model of Algorithmic Progress \citep{jones1995rd}}

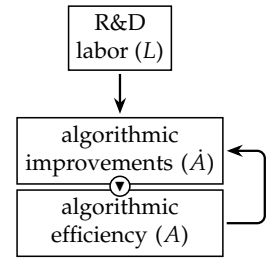
\begin{wrapfigure}{r}{0.5\textwidth}
\captionof{figure}{Jones Model of Innovation}\label{fig:basic-innovation}
\centering
\scalebox{\rsiscale}{%
\begin{tikzpicture}[rsi diagram, >={Stealth[]}]
    \rsiNode[align=center, thick, at={(0,0)}]{research}{R\&D\\labor ($L$)}
    \rsiNode[align=center, thick, minimum width=3.7cm, text width=3.5cm, at={(0,-3.4)}]{algorithms}{algorithmic\\efficiency ($A$)}
    \rsiNode[align=center, thick, minimum width=3.7cm, text width=3.5cm, anchor=south, at={([yshift=-0.8pt]algorithms.north)}]{adot}{algorithmic\\improvements ($\dot A$)}
    \draw[black, very thick, cg edge] (research.south) -- (adot.north);
    \draw[black, very thick, cg edge] (algorithms.east) -- ++(0.7,0) |- (adot.east);
    \node[draw, thick, circle, fill=white, inner sep=1pt] at (algorithms.north) {\scriptsize$\blacktriangledown$};
\end{tikzpicture}
}
\end{wrapfigure}

We start by applying the \citet{jones1995rd} model of innovation to AI progress.

We let $A$ denote algorithmic efficiency, defined as the inverse of the training compute required to reach a given level of capabilities.\footnote{One simple operationalization is the training time required to reach the GPT-2 level pretraining loss. Algorithmic progress has caused this to fall by a factor of 700 over 2019-2026, implying $\frac{A_{2026}}{A_{2019}}=700$. A fuller accounting of algorithmic progress would also account for downstream capabilities, and scale-dependence. We discuss these further below.}

Improvements to algorithmic efficiency $(\dot{A})$ are determined by two factors:
\begin{enumerate}
    \item The quantity of R\&D labor $(L)$.
    \item The current level of algorithmic efficiency ($A$). Higher achievements could either make improvements harder (diminishing returns) or easier (increasing returns).
\end{enumerate}

The second effect implies a \textit{feedback loop} between the stock of technology ($A$) and new technological improvements ($\adot{A}$). Its strength is characterized by the \textit{elasticity} of $\adot{A}$ to $A$, which is denoted $\varepsilon_{\smash{\adot{A}},A}$. 

\vspace{-0.5em}
\paragraph{Self-sustaining acceleration.} We say there is \textit{self-sustaining acceleration} in $A$ if, holding exogenous inputs fixed, the growth rate of technology ($\dot{A}/A$) is higher when the level of technology ($A$) is higher. In this model, the exogenous input held fixed is R\&D labor ($L$).

In this model, a self-sustaining acceleration occurs when the elasticity $\varepsilon_{\smash{\adot{A}},A} > 1$, meaning that a 1\% increase in the stock of ideas ($A$) causes an increase in the discovery of new ideas $(\adot{A})$ of more than 1\%, holding labor ($L$) constant. Empirically, in traditional industries ranging from agriculture to semiconductors, the elasticity is well below 1, implying stable growth of $A$ requires an ever-increasing quantity of R\&D effort \citep{bloom2020ideas}.  Studies examining historical elasticities of software and AI algorithmic progress also estimate that this condition does not hold without additional feedback loops \citep{erdil2024estimating,eth2025software}, which we discuss next.

\begin{tcolorbox}[
breakable,
  colback=newblue!5,
  colframe=newblue!100,
  title=Definition: Elasticity,
  fontupper=\fontsize{12pt}{14pt}\selectfont,
  fonttitle=\bfseries
]

An elasticity measures responsiveness, in proportional terms: the percentage change in an output caused by a one-percent change in an input, holding other inputs fixed. For an output variable $Y = f(X_1,\dots,X_n)$, the elasticity of $Y$ with respect to one of its inputs $X_j$ is
\[
\varepsilon_{Y,X_j} \;\equiv\; \frac{\partial \log Y}{\partial \log X_j}
\;=\; \frac{\partial f}{\partial X_j}  \,\frac{X_j}{Y}.
\]
Equivalently, it is the local slope of $Y$ against $X_j$ on logarithmic axes. For a power function $Y = X^{a}$, the elasticity is constant and equal to the exponent $a$.
\end{tcolorbox}

\begin{tcolorbox}[
breakable,
  colback=black!5,      
  colframe=black!70, 
  title=Technical Details: Graphs \& self-sustaining acceleration in technology,
  fontupper=\fontsize{12pt}{14pt}\selectfont,
  fonttitle=\bfseries
]
\paragraph{What do graph edges mean?} Graph edges ($\longrightarrow$) indicate that one variable is an input to the production function of another. The diagram for the Jones model implies there exists some function $\dot A = f(A,L)$. 

\vspace{1em}

The strength of a graph edge is determined by the elasticity of a variable with respect to its parents. Denote the two elasticities in the Jones model as:
\[
\varepsilon_{\smash{\adot{A}},A} = \frac{\partial \log f(A,L)}{\partial \log A} \quad \text{and} \quad 
\varepsilon_{\smash{\adot{A}},L} = \frac{\partial \log f(A,L)}{\partial \log L} 
\]
where $\varepsilon_{\smash{\adot{A}}, A}$ is the rate of diminishing returns to R\&D and $\varepsilon_{\smash{\adot{A}}, L}$ is the returns to scale in contemporaneous research effort. The Jones model is often written assuming that these two elasticities are constant ($\adot A=L^\lambda A^{1-\beta}$ with constant $\lambda$ and $\beta$), but we do not make that assumption.

\vspace{1em}

\paragraph{Accumulation.} Our directed graphs distinguish stock variables from flow variables to ensure that the diagrams map one-to-one to a system of equations without ambiguity \citep{richardson1986problems}. 
We use solid arrows ($\longrightarrow$) to represent how variables relate to each other at each point in time, and use triangles ($\blacktriangleright$) to denote how the rate of change of a stock variable, which is a flow (e.g. $\dot A$), accumulates into itself (e.g. into $A$) over time.

\vspace{1em}

\paragraph{Feedback loops and their strength.} In the Jones model, there is only one feedback loop, between technological improvements $\dot{A}$ and the level of technology $A$. Below we will consider models with many feedback loops, and ones where each feedback loop may traverse many edges ($A \longrightarrow B \longrightarrow C \longrightarrow \cdots$). Formally, each feedback loop is an ordered sequence of directed edges $\pmb{e} = (e_1,\ldots e_i,\ldots)$, whose strength is governed by the partial elasticities that we denote $\varepsilon_{e_i}$.

\vspace{1em}

The strength of any individual feedback loop $\pmb{e}$ is determined by the \textit{product of the elasticities along the edges}, $\Pi_i \varepsilon_{e_i}$.

\vspace{1em}

The strength of all feedback loops is determined by the \textit{total} elasticity of $\dot A$ with respect to $A$, which sums over all individual feedback loops:
\begin{align*}
\mathcal{E}_{\smash{\adot{A}},A}\ \equiv\ \frac{\mathrm d\log \dot A}{\mathrm d\log A}=\Bigg(\sum_{\substack{\text{paths } \pmb{e} \\ \text{from }A \text{ to }\dot A}}\ \prod_{\substack{i}}\varepsilon_{e_i}\Bigg)
\end{align*}
In Jones, with a single edge, the total elasticity is the partial elasticity: $\mathcal{E}_{\smash{\adot{A}}, A} = \varepsilon_{\smash{\adot{A}},A}$.

\vspace{1em}

\paragraph{Self-sustaining acceleration in technological growth.} If all exogenous inputs are held constant, the total elasticity determines how the growth rate of technology $g_A = \dot A / A$ changes as $A$ grows, yielding three regimes:
\[
\left\{
\begin{alignedat}{3}
\mathcal{E}_{\smash{\adot{A}},A} &< 1 &:{}\quad&
g_A \text{ falls as } A \text{ grows}
\quad&\Rightarrow{}\quad& \text{sub-exponential growth (fizzling out),}
\\
\mathcal{E}_{\smash{\adot{A}},A} &= 1 &:{}\quad&
g_A \text{ constant}
\quad&\Rightarrow{}\quad& \text{exponential growth (knife-edge),}
\\
\mathcal{E}_{\smash{\adot{A}},A} &> 1 &:{}\quad&
g_A \text{ rises as } A \text{ grows}
&\Rightarrow{}\quad& \text{self-sustaining acceleration.}
\end{alignedat}
\right.
\]

That is, $A$ shows self-sustaining acceleration when
\begin{align}
    \mathcal{E}_{\smash{\adot{A}},A} = \frac{\mathrm d\log\dot A}{\mathrm d\log A} > 1.
    \label{ssa:a}
\end{align}
In the Jones model, since there is a single feedback path $A\to \dot A$ with a single edge, growth shows self-sustaining acceleration when $\varepsilon_{\smash{\adot{A}},A} > 1$.

\vspace{1em}

\textbf{Non-constant elasticities and growth spurts.} Self-sustaining acceleration at one point in time does not imply self-sustaining acceleration forever, because the elasticities are not necessarily constant. In the bottlenecks section below we discuss this likely case. Even locally self-sustaining acceleration would be notable.

\end{tcolorbox}

\needspace{3.6in}
\subsection{Baseline Model of RSI}
\label{sec:simplemodel}

\begin{wrapfigure}{r}{0.5\textwidth}
\captionof{figure}{Baseline model of RSI}
\centering
\scalebox{\rsiscale}{%
\begin{tikzpicture}[rsi diagram, >={Stealth[]}]
    \rsiNode[align=center, thick, at={(0,-2.2)}]{erd}{R\&D\\ labor ($L$)}
    \rsiNode[align=center, thick, minimum width=3.7cm, text width=3.5cm, at={(0,-5.5)}]{algorithms}{algorithmic\\ efficiency ($A$)}
    \rsiNode[align=center, thick, minimum width=3.7cm, text width=3.5cm, anchor=south, at={([yshift=-0.8pt]algorithms.north)}]{adot}{algorithmic\\improvements ($\dot A$)}
    \rsiNode[draw=newblue, text=newblue, very thick, align=center, at={(0,-7.7)}]{capabilities}{capabilities\\($C$)}
    \rsiNode[draw=newblue, text=newblue, very thick, align=center, at={(-3.6,-7.7)}]{compute}{training\\compute ($T$)}
    \draw[black, very thick, cg edge] (erd.south) -- (adot.north);
    \draw[newblue, very thick, cg edge] (algorithms.south) -- (capabilities.north);
    \draw[newblue, very thick, cg edge] (compute.east) -- (capabilities.west);
    \draw[black, very thick, cg edge] (algorithms.east) -- (2.4,-5.5) |- (adot.east);
    \draw[newblue, very thick, cg edge] (capabilities.east) -- (2.9,-7.7) |- (adot.east);
    \node[draw, thick, circle, fill=white, inner sep=1pt] at (algorithms.north) {\scriptsize${\blacktriangledown}$};
\end{tikzpicture}
}
\end{wrapfigure}
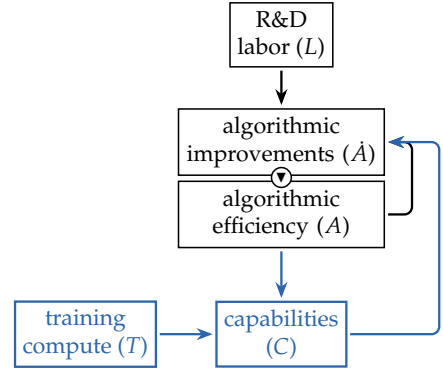

Our baseline model of RSI adds a few elements.
\begin{enumerate}[
    label=\arabic*.,
    leftmargin=2em,
    labelsep=0.8em,
    itemsep=0.2em,
]
    \item We introduce AI capabilities ($C$), which are determined by algorithmic efficiency ($A$) as well as training compute ($T$). Compute is held fixed, for now. 
    Measuring AI capabilities is notoriously difficult, but two common metrics are METR's time horizon measure (\cite{kwa2025horizon}) and the Epoch Capabilities Index (ECI, \cite{ho2025rosetta}), though both metrics have flaws.
    \item Algorithms are improved ($\adot{A}$) not just by human R\&D labor ($L$) but also with AI capabilities.
\end{enumerate}

Critically, the fact that AI capabilities now feed into algorithmic improvements creates a new feedback loop: the \textit{core feedback loop} ($A \to C \to \adot{A}$). This loop is in addition to a self-feedback loop ($A \to \adot{A}$) that appeared in the Jones model.

The strength of a multi-step feedback loop like the core loop is determined by the strength of all edges along it. In particular, the strength is the product of all elasticities along the loop, $\varepsilon_{\smash{\adot{A}},C}\varepsilon_{C,A}$. 

\paragraph{Self-sustaining acceleration.} Self-sustaining acceleration in algorithmic efficiency now depends on the combined strength of these two feedback loops. The condition is that the \textit{total} elasticity of $\dot A$ with respect to $A$, denoted $\mathcal{E}_{\smash{\adot{A}}, A}$, exceeds one:
\begin{align}
\mathcal{E}_{\smash{\adot{A}},A}
= \underbrace{\varepsilon_{\smash{\adot{A}},A}}_{\text{self-feedback loop}}
+ \underbrace{\varepsilon_{\smash{\adot{A}},C}\varepsilon_{C,A}}_{\text{core feedback loop}}
\; > \; 1
\label{core_feedback_loop_condition}
\end{align}

\paragraph{Self-sustaining acceleration in AI capabilities.} It is possible for algorithmic efficiency ($A$) to accelerate while capabilities ($C$) do not, if the elasticity from $A$ to $C$ is falling, for example if there was a hard ceiling on model capabilities. We discuss the issue further in Section~\ref{sec:narrow-broad-capabilities}.

\needspace{3.9in}
\subsection{Bottlenecks: Human, Compute, or Data} \label{sec:bottlenecks}

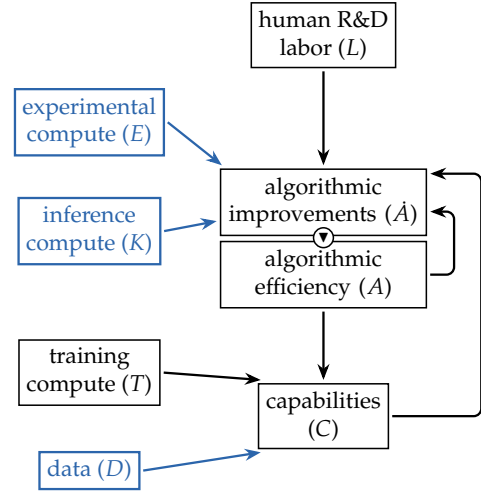
\begin{wrapfigure}{r}{0.5\textwidth}
\captionof{figure}{Model with bottlenecks}\label{fig:experiment-bottleneck}
\centering
\scalebox{\rsiscale}{%
\begin{tikzpicture}[rsi diagram, >={Stealth[]}]
    \rsiNode[align=center, thick, at={(0,0)}]{research}{human R\&D\\labor ($L$)}
    \rsiNode[align=center, thick, minimum width=3.7cm, text width=3.5cm, at={(0,-4.4)}]{algorithms}{algorithmic\\efficiency ($A$)}
    \rsiNode[align=center, thick, minimum width=3.7cm, text width=3.5cm, anchor=south, at={([yshift=-0.8pt]algorithms.north)}]{adot}{algorithmic\\improvements ($\dot A$)}
    \rsiNode[align=center, thick, at={(0,-7.0)}]{capabilities}{capabilities\\($C$)}
    \rsiNode[draw=newblue, text=newblue, very thick, align=center, at={(-4.3,-1.6)}]{experiments}{experimental\\compute ($E$)}
    \rsiNode[draw=newblue, text=newblue, very thick, align=center, at={(-4.3,-3.6)}]{inference}{inference\\compute ($K$)}
    \rsiNode[draw=black, thick, align=center, at={(-4.3,-6.2)}]{compute}{training\\compute ($T$)}
    \rsiNode[draw=newblue, text=newblue, very thick, align=center, at={(-4.3,-8.0)}]{data}{data ($D$)}
    \draw[black, very thick, cg edge] (research.south) -- (adot.north);
    \draw[black, very thick, cg edge] (algorithms.south) -- (capabilities.north);
    \draw[newblue, very thick, cg edge] (experiments.east) -- (adot.160);
    \draw[newblue, very thick, cg edge] (inference.east) -- (adot.190);
    \draw[black, very thick, cg edge] (compute.east) -- (capabilities.150);
    \draw[newblue, very thick, cg edge] (data.east) -- (capabilities.210);
    \draw[black, very thick, cg edge] (algorithms.east) -- (2.4,-4.4) |- (adot.-5);
    \draw[black, very thick, cg edge] (capabilities.east) -- (2.9,-7.0) |- (adot.15);
    \node[draw, thick, circle, fill=white, inner sep=1pt] at (algorithms.north) {\scriptsize${\blacktriangledown}$};
\end{tikzpicture}
}
\end{wrapfigure}

The elasticities in our baseline model may fall over time: the feedback loops could become
bottlenecked by critical inputs. Whether or not one input may bottleneck an output depends on how \textit{complementary} different inputs are.
\marginpar{\tiny TC \& BH resolved: we will remove R from this section.}

To emphasize the possibility of bottlenecks, we modify the baseline model to allow for three more inputs in the system. Experimental compute ($E$) and inference compute ($K$) are both used to make algorithmic improvements. Data ($D$), meanwhile, is used to produce AI capabilities.

Because the two feedback loops remain the same ($A \to \dot A$ and $A \to C \to \dot A$), the total elasticity remains the same as above, and thus the self-sustaining condition also remains the same as \eqref{core_feedback_loop_condition}. However these additional terms give us additional reasons why the core elasticity, $\varepsilon_{\adot{A}, C}$, may fall: the arrow from $C \to \dot A$ may weaken. We discuss these bottlenecks in turn:



\vspace{-0.5em}

\begin{itemize}
    \item \textbf{Labor.} Algorithmic improvements could become bottlenecked by a shortage of human labor if AI agents can only help with some but not all parts of the research process (e.g. due to ``lack of taste''). If inference compute increases disproportionately relative to human labor, bottlenecks reduce the marginal value of such compute in research \citep{jones2025artificial}.
    \item \textbf{Experimental compute.} AI labs often allocate a large part of their compute to experimentation. If model capabilities and experimental compute are strong complements, then this could be a bottleneck \citep{whitfill2025compute, ho2025experiments}.
    \item \textbf{Inference compute.} There is clearly a strong interaction between model capabilities and inference compute in producing algorithmic improvements. If they are strong complements then slow growth in inference compute could serve as a bottleneck, diminishing the effect of capabilities on algorithmic improvements.
    \item \textbf{Data.} A lack of high-quality data could bottleneck the advance of capabilities \citep{farboodi2025data,erdil2025automate,millidge2025data}. On the other hand, \citet{ho2024algorithmic} argue that most algorithmic progress is ``data-augmenting'', implying that data will not become a bottleneck.
    \item \textbf{Training compute.} A shortage of compute for training could also bottleneck the advance of AI capabilities \citep{gundlach2025origin}.
\end{itemize}


\begin{tcolorbox}[
  colback=black!5,      
  colframe=black!70, 
  title=Technical Details: The $C \longrightarrow \dot{A}$ arrow,
  fontupper=\fontsize{12pt}{14pt}\selectfont,
  fonttitle=\bfseries
]

\paragraph{Aside on AI capabilities $\rightarrow$ algorithmic improvements}
Note that human researchers ($L$), AI capabilities ($C$) and experimental compute, $E$, combine to produce algorithmic progress. Previous literature, such as \citet{eth2025software} and \citet{whitfill2025compute}, has tried to parametrize this function. But our view is that it is not obvious what this function looks like because it depends on ``the returns to intelligence" and how humans and AI systems differ, both of which we have a poor understanding of. Luckily the only thing we need to know is the elasticity noted above and it may be possible to measure this elasticity without fully understanding AI capabilities. We discuss this further in the data section. 

\end{tcolorbox}

\subsection{Narrow and Broad Capabilities} \label{sec:narrow-broad-capabilities}

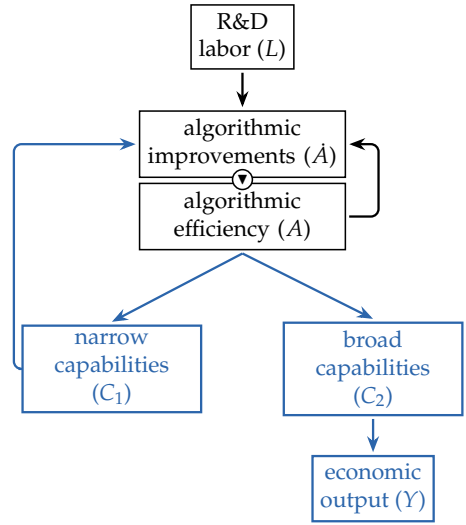
\begin{wrapfigure}{r}{0.5\textwidth}
    \captionof{figure}{Narrow capabilities feed back; broad capabilities dangle.}
    \centering
    \scalebox{\rsiscale}{%
    \begin{tikzpicture}[rsi diagram, >={Stealth[]}]
        \rsiNode[align=center, thick, at={(0,0)}]{labor}{R\&D\\labor ($L$)}
        \rsiNode[align=center, thick, minimum width=3.7cm, text width=3.5cm, at={(0,-3.3)}]{algorithms}{algorithmic\\efficiency ($A$)}
        \rsiNode[align=center, thick, minimum width=3.7cm, text width=3.5cm, anchor=south, at={([yshift=-0.8pt]algorithms.north)}]{adot}{algorithmic\\improvements ($\dot A$)}
        \rsiNode[draw=newblue, text=newblue, very thick, align=center, minimum width=3.2cm, text width=3.0cm, at={(-2.4,-6.1)}]{narrow}{narrow\\capabilities\\($C_1$)}
        \rsiNode[draw=newblue, text=newblue, very thick, align=center, minimum width=3.2cm, text width=3.0cm, at={(2.4,-6.1)}]{broad}{broad\\capabilities\\($C_2$)}
        \rsiNode[draw=newblue, text=newblue, very thick, align=center, at={(2.4,-8.3)}]{output}{economic\\output ($Y$)}
        \draw[black, very thick, cg edge] (labor.south) -- (adot.north);
        \draw[black, very thick, cg edge] (algorithms.east) -- (2.5,-3.3) |- (adot.east);
        \draw[newblue, very thick, cg edge] (algorithms.south) -- (narrow.north);
        \draw[newblue, very thick, cg edge] (algorithms.south) -- (broad.north);
        \draw[newblue, very thick, cg edge] (narrow.west) -- (-4.2,-6.1) |- (adot.west);
        \draw[newblue, very thick, cg edge] (broad.south) -- (output.north);
        \node[draw, thick, circle, fill=white, inner sep=1pt] at (algorithms.north) {\scriptsize${\blacktriangledown}$};
    \end{tikzpicture}
    }
\end{wrapfigure}

Our baseline model treats AI capabilities as a scalar $C$, which could be interpreted as the capability to both (1) contribute to algorithmic improvements ($\adot{A}$) and (2) contribute to producing economic output ($Y$). However, the core feedback loop only requires the $C \to \dot A$ arrow, not any arrow from $C \to Y$.

Here, we make explicit the possibility of a \textit{narrow} acceleration, where algorithmic efficiency accelerates but economic output does not. (For simplicity, this subsection suppresses training compute $T$.) The figure illustrates the model: we distinguish between a `narrow' capability ($C_1$) that is useful for contributing to algorithmic improvements, and a `broad' capability ($C_2$) that affects economic output ($Y$).

The clean cut we draw between capabilities useful for algorithmic improvements and all other capabilities is a strong simplification. In reality, it is plausible that a ``narrow" acceleration occurs across easier-to-verify tasks which includes optimizing algorithms but also many parts of mathematics, coding writ large, and more. Such progress would have impacts on the real economy but would be exceptionally jagged. It is also of course possible for progress in narrow tasks to eventually spill over into harder-to-verify tasks that matter for the broader economy.

\paragraph{Self-sustaining acceleration.} The condition for a self-sustaining acceleration in algorithmic efficiency remains as before, but with narrow capabilities $C_1$ substituted for $C$:
\[
\mathcal{E}_{\smash{\adot{A}},A}
= \underbrace{\varepsilon_{\smash{\adot{A}},A}}_{\text{self-feedback loop}}
+ \underbrace{\varepsilon_{\smash{\adot{A}},C_1}\varepsilon_{C_1,A}}_{\text{narrow capability loop}}
\]
Broad capabilities $C_2$ do not appear in this expression.

Such an acceleration in algorithmic efficiency $A$ need not imply a self-sustaining acceleration in broad capabilities $C_2$, and therefore need not imply such an acceleration in economic growth ($Y$). The condition for broad capabilities depends additionally on a \textit{superelasticity}: how the passthrough from algorithmic efficiency $A$ into broad capabilities $C_2$ changes as $A$ rises:\footnote{A corresponding condition for acceleration in output depends on an additional superelasticity from $C_2$ to $Y$; but for practical purposes we could define broad capabilities by their economic value, so $Y\propto C_2$.}
\begin{align}
    \varepsilon_{\smash{\adot{A}},A}
    + \varepsilon_{\smash{\adot{A}},C_1}\varepsilon_{C_1,A}
    + \underbrace{\frac{\mathrm d\log \varepsilon_{C_2,A}}{\mathrm d\log A}}_{\text{superelasticity}}
    > 1
\label{ssa:a_c}
\end{align}

Intuitively, the elasticity $\varepsilon_{C_2,A}$ could fall -- the superelasticity could be negative -- if traditional scaling was sufficient to surpass human abilities at algorithmic optimization ($C_1$), but not to surpass human abilities in other practical areas ($C_2$) due to unmodeled data bottlenecks on the latter.\footnote{This seems plausible because computers already outperform humans at many well-defined optimization tasks.}


\needspace{4.2in}
\subsection{Specific Optimization Ability and Growth Spurts} \label{sec:spurts}

\begin{wrapfigure}{r}{0.5\textwidth}
    \captionof{figure}{AI helps one subalgorithm}
    \centering
    \scalebox{\rsiscale}{%
    \begin{tikzpicture}[rsi diagram, >={Stealth[]}]
        \rsiNode[align=center, thick, at={(-2.8,0.9)}]{labor1}{human R\&D\\labor ($L_1$)}
        \rsiNode[align=center, thick, at={(2.8,0.9)}]{labor2}{human R\&D\\labor ($L_2$)}
        \rsiNode[align=center, thick, minimum width=3.4cm, text width=3.2cm, at={(-2.8,-3.4)}]{a1}{algorithm 1\\efficiency\\($A_1$)}
        \rsiNode[align=center, thick, minimum width=3.4cm, text width=3.2cm, anchor=south, at={([yshift=-0.8pt]a1.north)}]{a1dot}{improvements\\to algorithm 1\\($\dot A_1$)}
        \rsiNode[draw=newblue, text=newblue, very thick, align=center, minimum width=3.4cm, text width=3.2cm, at={(2.8,-3.4)}]{a2}{algorithm 2\\efficiency\\($A_2$)}
        \rsiNode[draw=newblue, text=newblue, very thick, align=center, minimum width=3.4cm, text width=3.2cm, anchor=south, at={([yshift=-0.8pt]a2.north)}]{a2dot}{improvements\\to algorithm 2\\($\dot A_2$)}
        \rsiNode[align=center, thick, minimum width=3.4cm, text width=3.2cm, at={(0,-6.6)}]{a}{overall\\efficiency\\($A$)}
        \rsiNode[align=center, thick, at={(0,-8.8)}]{cap}{capabilities\\($C$)}
        \draw[black, very thick, cg edge] (labor1.south) -- (a1dot.north);
        \draw[black, very thick, cg edge] (labor2.south) -- (a2dot.north);
        \draw[black, very thick, cg edge] (a1.east) -- ++(0.45,0) |- (a1dot.east);
        \draw[black, very thick, cg edge] (a2.west) -- ++(-0.45,0) |- (a2dot.west);
        \draw[black, very thick, cg edge] (a1.south) -- (a.150);
        \draw[black, very thick, cg edge] (a2.south) -- (a.30);
        \draw[black, very thick, cg edge] (a.south) -- (cap.north);
        \draw[newblue, very thick, cg edge] (cap.east) -- (5.6,-8.8) |- (a2dot.east);
        \node[draw, thick, circle, fill=white, inner sep=1pt] at (a1.north) {\scriptsize${\blacktriangledown}$};
        \node[draw, thick, circle, fill=white, inner sep=1pt] at (a2.north) {\scriptsize${\blacktriangledown}$};
    \end{tikzpicture}
    }
\end{wrapfigure}
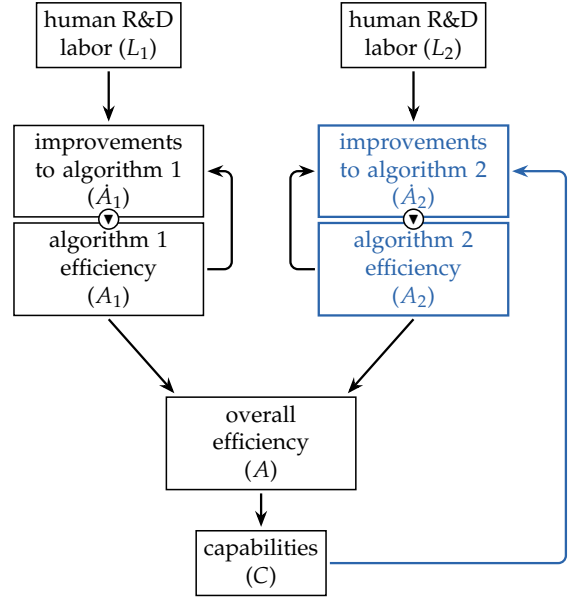

It seems very likely that AI will be relatively better at contributing to some parts of model training than others. This in turn is likely to cause ``growth spurts'' in AI progress.

The model at right assumes overall efficiency ($A$) depends on the efficiency of two sub-algorithms ($A_1$ and $A_2$), but AI directly helps improve only the second one, $C\rightarrow \adot{A}_2$. To isolate this AI-assisted channel, hold $A_1$ fixed. The relevant feedback loop is then $A_2\to A\to C\to\adot{A}_2$. Because $g_A=\varepsilon_{A,A_2}g_{A_2}$, self-sustaining acceleration in overall efficiency requires:
\[
    \underbrace{\varepsilon_{\smash{\adot{A}}_2,A_2}
    +\varepsilon_{\smash{\adot{A}}_2,C}\varepsilon_{C,A}\varepsilon_{A,A_2}}_{
    \mathrm d\log\dot A_2/\mathrm d\log A_2}
    +\underbrace{\frac{\mathrm d\log\varepsilon_{A,A_2}}{\mathrm d\log A_2}}_{\text{changing passthrough} \atop \text{into overall efficiency}}
    \ >\ 1.
\]

If these elasticities were constant then the second term would be zero, and self-sustaining acceleration today would imply self-sustaining acceleration forever. However there are two reasons why the condition may cease to hold, producing only a temporary ``growth spurt'':

\begin{enumerate}
    \item \textbf{Parallel algorithms.} Overall algorithmic efficiency could be bottlenecked by type-1 algorithmic efficiency, which is not improved by AI capabilities. As type-2 algorithms become abundant, the arrow from $A_2 \to A$ weakens: $\varepsilon_{A,A_2}$ falls and its log derivative is negative. For example, if the compute costs of two necessary processes add, so that $1/A=1/A_1+1/A_2$, then as $A_2/A_1\to\infty$,
    \[
        \varepsilon_{A,A_2}\to0
        \qquad\text{and}\qquad
        \frac{\mathrm d\log\varepsilon_{A,A_2}}{\mathrm d\log A_2}\to-1.
    \]
    Thus a spurt driven by progress on the easy component stalls unless AI also starts accelerating progress on the bottleneck component.
    \item \textbf{Low ceilings.} If one of the algorithms is already close to a natural ceiling then its elasticity must necessarily fall. Some algorithms are naturally limited, e.g. if kernel efficiency is currently 50\%, it can only increase by a factor of two; while the efficiency of other algorithms has improved by orders of magnitude, and may continue to improve by large factors.
\end{enumerate}

Empirically, it will be useful to document which types of algorithms are receiving the biggest boost from AI-assisted research. On first principles, we might expect bigger boosts where there are low costs of verification (e.g., optimizing inference, optimizing elicitation) than where there are high costs of verification (optimizing architectures). Additionally, it is plausible that improvements to type-1 algorithms do advance capabilities, but only `narrowly' in the spirit of the model in Section \ref{sec:narrow-broad-capabilities}. We leave combining the two models for future work.

\needspace{4.2in}
\subsection{Economic Feedback Loops} \label{sec:economicfeedback}

\begin{wrapfigure}{r}{0.5\textwidth}
\captionof{figure}{Model with Economic Feedback Loops}
\label{fig:full-system}
\centering
\scalebox{\rsiscale}{%
\begin{tikzpicture}[rsi diagram, >={Stealth[]}]

    \rsiNode[
        align=center,
        thick,
        at={(0,-1.3)}
    ]{research}{human R\&D\\labor ($L$)}

    \rsiNode[
        align=center,
        thick,
        minimum width=3.7cm,
        text width=3.5cm,
        at={(0,-5.5)}
    ]{algorithms}{algorithmic\\efficiency ($A$)}

    \rsiNode[
        align=center,
        thick,
        minimum width=3.7cm,
        text width=3.5cm,
        anchor=south,
        at={([yshift=-0.8pt]algorithms.north)}
    ]{adot}{algorithmic\\improvements ($\dot A$)}

    \rsiNode[
        align=center,
        thick,
        at={(0,-7.7)}
    ]{capabilities}{capabilities\\($C$)}

    \rsiNode[
        draw=newblue,
        text=newblue,
        very thick,
        align=center,
        at={(0,-9.9)}
    ]{output}{economic\\output ($Y$)}

    \rsiNode[
        align=center,
        thick,
        minimum height=1.1cm,
        at={(-4.3,-2.1)}
    ]{experiments}{experimental\\compute ($E$)}

    \rsiNode[
        align=center,
        thick,
        minimum height=1.1cm,
        at={(-4.3,-4.3)}
    ]{inference}{inference\\compute ($K$)}

    \rsiNode[
        align=center,
        thick,
        minimum height=1.1cm,
        at={(-4.3,-6.6)}
    ]{compute}{training\\compute ($T$)}

    \rsiNode[
        align=center,
        thick,
        minimum height=1.1cm,
        at={(-4.3,-8.8)}
    ]{data}{\;data ($D$)\;}

    \rsiNode[
        draw=newblue,
        text=newblue,
        very thick,
        align=center,
        minimum width=0.8cm,
        minimum height=1.1cm,
        anchor=east,
        at={([xshift=0.8pt]experiments.west)}
    ]{edot}{$\dot{E}$}

    \rsiNode[
        draw=newblue,
        text=newblue,
        very thick,
        align=center,
        minimum width=0.8cm,
        minimum height=1.1cm,
        anchor=east,
        at={([xshift=0.8pt]inference.west)}
    ]{kdot}{$\dot{K}$}

    \rsiNode[
        draw=newblue,
        text=newblue,
        very thick,
        align=center,
        minimum width=0.8cm,
        minimum height=1.1cm,
        anchor=east,
        at={([xshift=0.8pt]compute.west)}
    ]{tdot}{$\dot{T}$}

    \rsiNode[
        draw=newblue,
        text=newblue,
        very thick,
        align=center,
        minimum width=0.8cm,
        minimum height=1.1cm,
        anchor=east,
        at={([xshift=0.8pt]data.west)}
    ]{ddot}{$\dot{D}$}

    \draw[black, very thick, cg edge]
        (research.south) -- (adot.north);

    \draw[black, very thick, cg edge]
        (algorithms.south) -- (capabilities.north);

    \draw[newblue, very thick, cg edge]
        (capabilities.south) -- (output.north);

    \draw[black, very thick, cg edge]
        (experiments.east) -- (adot.160);

    \draw[black, very thick, cg edge]
        (inference.east) -- (adot.190);

    \draw[black, very thick, cg edge]
        (compute.east) -- (capabilities.150);

    \draw[black, very thick, cg edge]
        (data.east) -- (capabilities.210);

    \draw[black, very thick, cg edge]
        (algorithms.east) -- (2.4,-5.5) |- (adot.-10);

    \draw[black, very thick, cg edge]
        (capabilities.east) -- (2.9,-7.7) |- (adot.10);

    \draw[newblue, very thick, cg edge]
        (output.west) -- (-7.2,-9.9) -- (-7.2,-2.1) -- (edot.west);

    \draw[newblue, very thick, cg edge]
        (output.west) -- (-7.2,-9.9) -- (-7.2,-4.3) -- (kdot.west);

    \draw[newblue, very thick, cg edge]
        (output.west) -- (-7.2,-9.9) -- (-7.2,-6.6) -- (tdot.west);

    \draw[newblue, very thick, cg edge]
        (output.west) -- (-7.2,-9.9) -- (-7.2,-8.8) -- (ddot.west);

    \node[
        draw,
        thick,
        circle,
        fill=white,
        inner sep=1pt
    ] at (algorithms.north) {\scriptsize$\blacktriangledown$};

    \node[
        draw=newblue,
        text=newblue,
        thick,
        circle,
        fill=white,
        inner sep=1pt
    ] at (experiments.west) {\scriptsize$\blacktriangleright$};

    \node[
        draw=newblue,
        text=newblue,
        thick,
        circle,
        fill=white,
        inner sep=1pt
    ] at (inference.west) {\scriptsize$\blacktriangleright$};

    \node[
        draw=newblue,
        text=newblue,
        thick,
        circle,
        fill=white,
        inner sep=1pt
    ] at (compute.west) {\scriptsize$\blacktriangleright$};

    \node[
        draw=newblue,
        text=newblue,
        thick,
        circle,
        fill=white,
        inner sep=1pt
    ] at (data.west) {\scriptsize$\blacktriangleright$};

\end{tikzpicture}
}
\end{wrapfigure}
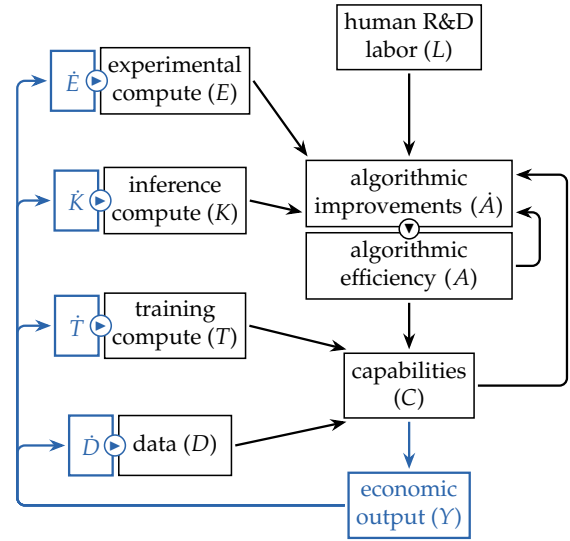
The previous models held constant the level of compute (for training, inference, or experiments) and held constant the quantity of data. This isolated the role of feedback loops driven by algorithmic improvements.

There are also economic feedback loops \citep{davidson2026automating}:
\begin{enumerate}
    \item Increased AI capabilities $(C)$ drive higher economic output $(Y)$.
    \item Higher economic output in turn finances further investment in compute (for experiments $E$, for inference $K$, and for training $T$), which feeds back into AI capabilities.
    \end{enumerate}
AI-induced economic growth also generates more data, likewise increasing AI capabilities, and further increasing economic growth \citep{farboodi2025data}.

We emphasize that economic output ($Y$) -- and the model as a whole -- can be interpreted macroeconomically, where $Y$ is GDP, or microeconomically, where $Y$ is the revenue of a single AI firm. In the latter case, as AI capabilities progress and the firm grows, more chips or data are affordable (without substantially altering factor prices). Such microeconomic effects are self-evidently already occurring, and the resulting loops may drive capability acceleration even without substantial macroeconomic impacts.

\paragraph{Feedback loops and self-sustaining acceleration.} Aside from the two feedback loops discussed above ($A \to \adot{A}$ and $A \to C \to \adot{A}$), self-sustaining acceleration in algorithmic efficiency now depends on the strength of additional economic feedback loops that run through AI capabilities. For instance, capabilities might drive the accumulation of high-quality data, which in turn drives capabilities ($C \to Y \to \adot{D} \to D \to C$), or the accumulation of training compute, which in turn drives capabilities ($C \to Y \to \adot{T} \to T \to C$).

The condition for a self-sustaining acceleration is now substantially more complicated and requires taking into account these more complicated feedback loops. The technical details box below states the condition; \citet{davidson2026automating} offers a general framework and solution.

\begin{tcolorbox}[
breakable,
  colback=black!5,      
  colframe=black!70, 
  title=Technical Details: Self-sustaining acceleration with economic feedback loops,
  fontupper=\fontsize{12pt}{14pt}\selectfont,
  fonttitle=\bfseries
]
The prior boxes on technical details apply only when there is a single accumulating state: before, this was merely $A$. Now, instead, there is an entire state vector:
\begin{align*}
    X \equiv (A, T, D, E, K)'
\end{align*}
Define the reduced-form elasticity matrix $M$, with $M_{ij} \equiv \frac{\mathrm d \log \dot{X}_i}{\mathrm d \log {X}_j}$. Each entry of $M$ measures how strongly one accumulable input increases the accumulation of another, after tracing through the contemporaneous graph. 

\vspace{1em}



\citet{davidson2026automating} show that there is self-sustaining acceleration in algorithmic efficiency when $\rho(M) > 1$, where $\rho(M)$ is the dominant eigenvalue of the feedback matrix $M$.

\vspace{1em}

After some math, it can be shown that the eigenvalue condition holds when the sum of three feedback channels exceeds one: the core feedback loop ($A \to C \to \adot{A}$); the indirect loop between economic output and capabilities through training compute and data; and the direct economic loop through experimental and inference compute 
for algorithmic improvements. 
In math:
\begin{align}
&
\underbrace{
    \frac{
        \varepsilon_{\smash{\dot A},C}\varepsilon_{C,A}
    }{
        1-\varepsilon_{\smash{\dot A},A}
    }
}_{\text{core feedback loop}}
\nonumber\\[0.5em]
&\qquad+
\underbrace{
    \varepsilon_{Y,C}
    \left(
        \varepsilon_{C,T}+\varepsilon_{C,D}
    \right)
}_{\text{indirect economic loops}}
\nonumber\\[0.5em]
&\qquad+
\underbrace{
    \frac{
        \varepsilon_{C,A}\varepsilon_{Y,C}
    }{
        1-\varepsilon_{\smash{\dot A},A}
    }
    \left(
        \varepsilon_{\smash{\dot A},E}
        +
        \varepsilon_{\smash{\dot A},K}
    \right)
}_{\substack{\text{direct economic loops through}\\ \text{experimental and inference compute}}}
>1.
\label{ssa:econ}
\end{align}

Note that this condition assumes that the direct algorithmic efficiency self-loop does not on its own power a self-sustaining acceleration ($\varepsilon_{\smash{\adot{A}},A} < 1$).\footnote{Analogously, the rearrangement into this three-term form assumes the indirect economic loop is not self-sustaining on its own, $\varepsilon_{Y,C}\left(\varepsilon_{C,T}+\varepsilon_{C,D}\right) < 1$; if it were, acceleration would occur through that loop alone.} It also imposes the Solow-like assumption that capital-like variables accumulate one-for-one with output: $\varepsilon_{\smash{\adot{T}},Y} = \varepsilon_{\smash{\adot{D}},Y} = \varepsilon_{\smash{\adot{E}},Y} = \varepsilon_{\smash{\adot{K}},Y} = 1$. If there are no economic impacts of AI capabilities ($\varepsilon_{Y,C} = 0$), the condition collapses to the earlier condition without economic feedback loops.

\vspace{1em}

A subtlety is that the second loop does not run through algorithmic efficiency $A$ and is self-contained: higher capabilities produce more output, financing more training compute and raising capabilities ($C \to Y \to \adot{T} \to T \to C$), or producing more data and raising capabilities ($C \to Y \to \adot{D} \to D \to C$). In either of these loops, the economy achieves a self-sustaining acceleration in algorithmic efficiency $A$ `merely' by raising GDP, which produces more compute or data \citep{davidson2026automating}.

\vspace{1em}

An additional takeaway is that the condition with economic feedback loops, \eqref{ssa:econ}, requires estimating five new elasticities: the elasticity of economic output to AI capabilities $\varepsilon_{Y,C}$; the elasticity of capabilities to training compute $\varepsilon_{C,T}$; the elasticity of capabilities to data $\varepsilon_{C,D}$; the elasticity of algorithmic improvements to experimental compute $\varepsilon_{\smash{\adot{A}},E}$; and the elasticity of algorithmic improvements to inference compute $\varepsilon_{\smash{\adot{A}},K}$.


\end{tcolorbox}

\endgroup

\ifSubfilesClassLoaded{%
    \paperbibliography%
}{}

\end{document}

\section{Data Relevant to RSI}\label{sec:data}

We primarily focus this section on calibrating the basic bottlenecks model. This is motivated primarily by simplicity and because larger economic feedback loops will be slower. 

To get empirical traction, we modify the bottlenecks model by assuming the existence of an ``R\&D effort aggregator ($R$)'' which combines human researchers ($L$), automated AI researchers (running on inference compute, $K$), and experimental compute ($E$) into a single scalar. R\&D effort is then the sole input to algorithmic improvements ($\adot{A}$). This essentially assumes that all R\&D inputs interact with the existing stock of algorithmic progress in the same way. This will let us use historical data on human inputs to calibrate what will happen with AI inputs. 

Once we make this assumption, we can expand and re-arrange the condition for self-sustaining acceleration in algorithmic efficiency,  \eqref{core_feedback_loop_condition}:\footnote{\label{fn:super} The condition \eqref{core_feedback_loop_condition} delineates a self-sustaining acceleration in algorithmic efficiency. In this section, we are most interested in a self-sustaining acceleration in \textit{AI capabilities}. The condition for capabilities is the same as the condition for efficiency, augmented with a fourth, `superelasticity' term, $\frac{1}{1-\varepsilon_{\adot A,A}} \frac{\mathrm d\log \varepsilon_{C,A}}{\mathrm d\log A}$. This measures the change in the capabilities elasticity $\varepsilon_{C,A}$ as algorithmic efficiency $A$ grows. If $\varepsilon_{C,A}$ is constant, the superelasticity term is zero and the condition reduces to the three terms shown here. In Section~\ref{subsec:superelasticity}, which we put in a technical box, we argue the evidence indeed points to a superelasticity near zero.}

\begin{align}
\underbrace{
\frac{\varepsilon_{\adot A,R}}{1-\varepsilon_{\adot A,A}}
}_{\text{return to R\&D}}
\cdot
\underbrace{
\varepsilon_{R,C}
}_{\substack{\text{research effort's} \\ \text{elasticity to} \\ \text{AI capabilities}}}
\cdot
\underbrace{
\varepsilon_{C,A}
}_{\substack{\text{capabilities'} \\ \text{elasticity to} \\ \text{algorithmic} \\ \text{efficiency}}}
\;>\;
1
\label{eq:measurement_condition}
\end{align}

\begin{tcolorbox}[
breakable,
  colback=black!5,
  colframe=black!70,
  title=Technical details: rearranging the self-sustaining acceleration condition,
  fontupper=\fontsize{12pt}{14pt}\selectfont,
  fonttitle=\bfseries
]
The condition for a self-sustaining acceleration in algorithmic efficiency, under the assumption of an R\&D effort aggregator $R$, is
\[
\varepsilon_{\adot A,A} + \varepsilon_{\adot A,R}\,\varepsilon_{R,C}\,\varepsilon_{C,A} 
\;>\; 1.
\]
This is the same as the core feedback loop condition used in the bottlenecks section, \eqref{core_feedback_loop_condition}, except that now AI capabilities $C$ pass through the $R$ aggregator to affect $\adot{A}$.

\vspace{1em}

Subtracting the direct-spillover term and dividing both sides by $1-\varepsilon_{\adot A,A}$ isolates the capability loop:
\[
\frac{\varepsilon_{\adot A,R}\,\varepsilon_{R,C}\,\varepsilon_{C,A}}{1-\varepsilon_{\adot A,A}}
\;>\; 1.
\]
Note that the denominator is positive whenever the direct spillover alone is not already self-sustaining, $\varepsilon_{\adot A,A}<1$. Then, grouping the first factor of the first term gives the form above,
\[
\underbrace{
\frac{\varepsilon_{\adot A,R}}{1-\varepsilon_{\adot A,A}}
}_{\text{return to R\&D}}
\cdot
\underbrace{
\varepsilon_{R,C}
}_{\substack{\text{research effort's} \\ \text{elasticity to} \\ \text{AI capabilities}}}
\cdot
\underbrace{
\varepsilon_{C,A}
}_{\substack{\text{capabilities'} \\ \text{elasticity to} \\ \text{algorithmic} \\ \text{efficiency}}}
\;>\;
1.
\]
\end{tcolorbox}

In our view, estimating this quantity, tracking it over time, and extrapolating when it might cross 1 is arguably the most important empirical exercise for assessing whether a self-sustaining acceleration will occur. The rest of this section maps each term in the condition to the data needed to estimate it: Section~\ref{subsec:research-to-ideas} covers the return to research, Section~\ref{subsec:research-to-capabilities} covers the elasticity of research effort with respect to AI capabilities, and Section~\ref{subsec:capabilities-to-algo} covers the response of capabilities to algorithmic progress. We then collect detailed supplementary evidence on model-training inputs and model economics to build a broader picture of AI R\&D automation.

\subsection{Return to Research} 
\label{subsec:research-to-ideas}

It turns out -- derived in the technical details box below -- that the first term, the ``returns to research'', can be measured from two growth rates: the growth rate of algorithmic efficiency ($g_A \equiv \dot{A}/A$) and the growth rate of R\&D effort ($g_R \equiv \dot{R}/R$). On a balanced growth path:
\[
\underbrace{
\frac{\varepsilon_{\adot A,R}}{1-\varepsilon_{\adot A,A}}
}_{\text{return to R\&D}}
= \frac{g_A}{g_R}
\]
We therefore need to estimate the growth rate of algorithmic efficiency, $g_A$, and the growth rate of research effort, $g_R$:
\begin{enumerate}
    \item We can measure growth in algorithmic efficiency $g_A$ directly from model-training data.
    \item We derive in a technical box below that one can measure R\&D effort as the \textit{spend-weighted average of growth rates in inputs}: human researchers $L$, experimental compute $E$, and inference compute used for research $K$. Using $S$ to denote spend shares,
    \begin{align*}
        g_R=\sum_{\text{input}} S_{\text{input}}\,g_{\text{input}}
    \end{align*}
    Alternatively, we can measure $g_R$ directly from spending. Total R\&D spend is the sum across inputs of price times quantity, and log-differentiating shows its growth splits into a price component and a quantity component:
    \begin{align*}
        g_{\text{spend}} = \underbrace{\sum_{\text{input}} S_{\text{input}}\, g_{\text{price of input}}}_{\equiv\, g_{\text{prices}} \text{, input price inflation}} + \underbrace{\sum_{\text{input}} S_{\text{input}}\, g_{\text{input}}}_{=\, g_R \text{, by the result above}}
    \end{align*}
    so $g_R = g_{\text{spend}} - g_{\text{prices}}$: growth in total R\&D spend, deflated by an index of R\&D input prices.
\end{enumerate}

\begin{longtable}[]{@{}
  >{\raggedright\arraybackslash}p{(\columnwidth - 4\tabcolsep) * \real{0.21}}
  >{\raggedright\arraybackslash}p{(\columnwidth - 4\tabcolsep) * \real{0.44}}
  >{\raggedright\arraybackslash}p{(\columnwidth - 4\tabcolsep) * \real{0.36}}@{}}
\toprule
\begin{minipage}[b]{\linewidth}\raggedright
\textbf{Model primitive}
\end{minipage} & \begin{minipage}[b]{\linewidth}\raggedright
\textbf{Existing data}
\end{minipage} & \begin{minipage}[b]{\linewidth}\raggedright
\textbf{Data ask}
\end{minipage} \\
\midrule
\endhead
\begin{minipage}[t]{\linewidth}\raggedright\setlength{\parskip}{1em}
\textbf{Growth in algorithmic efficiency ($g_A$)}

\end{minipage} & \begin{minipage}[t]{\linewidth}\raggedright\setlength{\parskip}{1em}
\citet{ho2024algorithmic} estimate pretraining algorithmic progress of 3x a year (i.e. compute needed to achieve a given level of capability is 1/3 as much each year). \citet{whitfill2025forecasting} estimate 3.5x/year for the entire stack. \citet{ho2026least} estimates 10x/year for the entire stack.

\citet{gundlach2025origin} argue that algorithmic progress has grown at different rates across scales: 1.5x/year at small scale, and 2.5x/year at frontier scale, for pretraining.
\end{minipage} & \begin{minipage}[t]{\linewidth}\raggedright\setlength{\parskip}{1em}
\textbf{Current data weakness}: Compute estimates for models since 2024 are uncertain.

\textbf{Ideal new data}: model training compute after 2024 to estimate $g_A$.
\end{minipage} \\
\addlinespace[1.2em]
\begin{minipage}[t]{\linewidth}\raggedright\setlength{\parskip}{1em}
\textbf{Share of human researchers ($L$) of R\&D spend, and its growth $g_L$}

Its share $S_L$ and growth $g_L$ enter $g_R=\sum_{\text{input}} S_{\text{input}}\,g_{\text{input}}$.
\end{minipage} & \begin{minipage}[t]{\linewidth}\raggedright\setlength{\parskip}{1em}
\citet{epoch2026aicompanies} estimates growth rates of around 2--3x/year in total staff for frontier labs over 2023--2025.

\citet{whitfill2025compute} estimate headcount growth rates of 2--3X/year for Anthropic and OpenAI over 2022--2024.

\citet{cottier2024cost} estimates that around 1/3 of the cost of training models was spent on researcher salaries.
\end{minipage} & \begin{minipage}[t]{\linewidth}\raggedright\setlength{\parskip}{1em}
\textbf{Current data weakness}: Employee headcount does not necessarily map to the researcher headcount. 

\textbf{Ideal new data}: Granular lab data on staff, split by seniority and by researchers vs.\ engineers, would let us treat these as separate inputs rather than an aggregate $L$.
\end{minipage} \\
\addlinespace[1.2em]
\begin{minipage}[t]{\linewidth}\raggedright\setlength{\parskip}{1em}
\textbf{Share of experimental compute ($E$) of R\&D spend, and its growth $g_E$}

Its share $S_E$ and growth $g_E$ enter $g_R=\sum_{\text{input}} S_{\text{input}}\,g_{\text{input}}$.
\end{minipage} & \begin{minipage}[t]{\linewidth}\raggedright\setlength{\parskip}{1em}
\citet{you2025openai} estimates that in 2024 only 10\% of OpenAI's R\&D compute was on final training runs, while 90\% was on experimental compute (i.e. training which did not directly affect weights used in a public product).

\citet{denain2026rnd} estimate similar ratios for Chinese companies.

\citet{morrison2026hidden} estimate similar ratios for OLMo 3 family.

\citet{epoch2026aichipproduction} finds that the total growth rate of chips is about $3.3x$ a year (doubling every 7 months). 

\end{minipage} & \begin{minipage}[t]{\linewidth}\raggedright\setlength{\parskip}{1em}
\textbf{Current data weakness}: Experimental compute share in total compute is not the same as the experimental compute share of total R\&D spending. The composition of experimental compute is also not clear. 

\textbf{Ideal new data}: Labs could publish detailed data on experimental compute spend across domains and over time, which would help estimate how much experimental compute bottlenecks automated research.
\end{minipage} \\
\addlinespace[1.2em]
\begin{minipage}[t]{\linewidth}\raggedright\setlength{\parskip}{1em}
\textbf{Share of inference compute ($K$) of R\&D spend, and its growth $g_K$}

Its share $S_K$ and growth $g_K$ enter $g_R=\sum_{\text{input}} S_{\text{input}}\,g_{\text{input}}$.
\end{minipage} & \begin{minipage}[t]{\linewidth}\raggedright\setlength{\parskip}{1em}
No data available.
\end{minipage} & \begin{minipage}[t]{\linewidth}\raggedright\setlength{\parskip}{1em}
\textbf{Ideal new data}: Labs could publish detailed data on inference compute spend across domains and over time.
\end{minipage} \\
\addlinespace[1.2em]
\begin{minipage}[t]{\linewidth}\raggedright\setlength{\parskip}{1em}
\textbf{Total R\&D spend in dollars, and its growth}

Deflated by an index of input prices, this is an alternative way to measure $g_R$.
\end{minipage} & \begin{minipage}[t]{\linewidth}\raggedright\setlength{\parskip}{1em}
\citet{denain2026rnd} estimate R\&D compute spending for several labs from public filings.
\end{minipage} & \begin{minipage}[t]{\linewidth}\raggedright\setlength{\parskip}{1em}
\textbf{Current data weakness}: No spending over time yet. Even given spend, no off-the-shelf deflator fits AI R\&D, since compute prices fall rapidly while wages do not.

\textbf{Ideal new data}: Labs could publish total annual R\&D expenditure alongside a self-constructed deflator. 
\end{minipage} \\
\addlinespace[1.2em]
\begin{minipage}[t]{\linewidth}\raggedright\setlength{\parskip}{1em}
\textbf{Substitutability of labor and experimental compute}

How substitutable labor and compute are controls how fast $\varepsilon_{R,C}$ falls as research becomes bottlenecked by experimental compute.
\end{minipage} & \begin{minipage}[t]{\linewidth}\raggedright\setlength{\parskip}{1em}
\citet{whitfill2025compute} try to estimate the substitutability between labor and experimental compute in AI R\&D, using variation in relative prices, but their findings are inconclusive.
\end{minipage} & \begin{minipage}[t]{\linewidth}\raggedright\setlength{\parskip}{1em}
\textbf{Current data weakness}: Aside from data quality, it is unclear whether the need for experimental compute increases with higher training compute.

\textbf{Ideal new data}: labs could conduct experiments that would give teams randomized amounts of R\&D labor and experimental compute, but this is expensive.

One cheaper substitute is using AI labor in place of human labor in the experiment.

Cheaper still is to survey managers how much more progress they would make with (A) twice the researchers (compute fixed); (B) twice the compute (researchers fixed); (C) both doubled.
\end{minipage} \\
\bottomrule
\end{longtable}

\begin{tcolorbox}[
breakable,
  colback=black!5,
  colframe=black!70,
  title=Technical details: cost minimization and elasticities,
  fontupper=\fontsize{12pt}{14pt}\selectfont,
  fonttitle=\bfseries
]
Throughout this section we use the correspondence that, under cost-minimizing input choice, the elasticity of an output with respect to an input equals that input's share of expenditure.

\paragraph{Setup.} Let output $R = R(X_1,\dots,X_n)$ be produced from inputs $X_i$ with prices $p_i$. A firm minimizing the cost $\sum_i p_i X_i$ of attaining a target $\bar R$ has first-order conditions $p_i = \mu\,\partial R/\partial X_i$ for a common multiplier $\mu$. The cost share of input $i$ is then
\[
S_i \equiv \frac{p_i X_i}{\sum_j p_j X_j}
= \frac{(\partial R/\partial X_i)\,X_i}{\sum_j (\partial R/\partial X_j)\,X_j}.
\]

\paragraph{Constant returns to scale.} We take $R$ to have constant returns to scale. This is without loss of generality: $R$ is a latent index, and its returns to scale are absorbed by the elasticity $\varepsilon_{\adot A,R}$ in the return-to-research term. Under constant returns, Euler's theorem gives $\sum_j (\partial R/\partial X_j)\,X_j = R$, so
\[
S_i = \frac{\partial R}{\partial X_i}\,\frac{X_i}{R} = \varepsilon_{R,X_i}.
\]
Cost shares equal elasticities.

\paragraph{When firms do not optimize.} If input choices are not fully cost-minimizing, the equality becomes a revealed-preference bound: observed cost shares reflect the firm's own belief about marginal products, so they remain informative about the elasticities.
\end{tcolorbox}

\begin{tcolorbox}[
breakable,
  colback=black!5,
  colframe=black!70,
  title=Technical details: return to research,
  fontupper=\fontsize{12pt}{14pt}\selectfont,
  fonttitle=\bfseries
]
\paragraph{The return to research as $g_A/g_R$.} Write the discovery rate as $\dot A = f(R,A)$, so $g_A \equiv \dot A/A = f(R,A)/A$. Log-differentiating with respect to time,
\[
\frac{\mathrm d\log g_A}{\mathrm d t} = \varepsilon_{\dot A,R}\,g_R - \big(1-\varepsilon_{\dot A,A}\big)\,g_A.
\]
On a balanced growth path $g_A$ is constant, so the left-hand side is zero and
\[
\frac{g_A}{g_R} = \frac{\varepsilon_{\dot A,R}}{1-\varepsilon_{\dot A,A}},
\]
which is exactly the return-to-research term. This holds exactly on a balanced growth path and approximately when $g_A$ is approximately constant. The smooth, steady historical rate of algorithmic progress makes this a reasonable benchmark, though not one we expect to hold if RSI pushes us off the balanced-growth path.

\vspace{1em}

\paragraph{Decomposing $g_R$.} The growth rate of research effort follows the growth-rate chain rule across its inputs,
\[
g_R = \sum_{\text{input } i} \varepsilon_{R,i}\, g_i.
\]
Using the cost-share correspondence above, $\varepsilon_{R,i}=S_i$, so
\[
g_R = \sum_{\text{input } i} S_i\, g_i,
\]
a spend-weighted average of input growth rates over human labor $L$, experimental compute $E$, and inference compute $K$.
\end{tcolorbox}

\subsection{Research Effort to Capabilities}\label{subsec:research-to-capabilities}

The second term measures how strongly effective research effort responds to more capable AIs. This is the hardest of the three to measure, since it captures how AI capabilities translate into real-world research effort. This requires scientifically understanding AI capabilities and how they feed into research. The science of AI capabilities is in its infancy.

One possible route is to measure it directly, through studies of how much AI accelerates end-to-end research. 

A second route uses a cost-share argument. If we model ``effective AI labor'' as a Cobb--Douglas function of AI capabilities and the quantity of AI researchers (inference compute), e.g.\ $C^\gamma K$, then this term equals $\gamma S_K$, where $\gamma$ is the returns to capabilities relative to scaling the quantity of AI labor and $S_K$ is the inference-compute cost share (derived in the technical details below).

\begin{longtable}[]{@{}
  >{\raggedright\arraybackslash}p{(\columnwidth - 4\tabcolsep) * \real{0.37}}
  >{\raggedright\arraybackslash}p{(\columnwidth - 4\tabcolsep) * \real{0.32}}
  >{\raggedright\arraybackslash}p{(\columnwidth - 4\tabcolsep) * \real{0.31}}@{}}
\toprule
\begin{minipage}[b]{\linewidth}\raggedright
\textbf{Model primitive}
\end{minipage} & \begin{minipage}[b]{\linewidth}\raggedright
\textbf{Existing data}
\end{minipage} & \begin{minipage}[b]{\linewidth}\raggedright
\textbf{Data ask}
\end{minipage} \\
\midrule
\endhead
\begin{minipage}[t]{\linewidth}\raggedright\setlength{\parskip}{1em}
\textbf{Share of AI R\&D inference compute ($K$) of total R\&D spend ($S_K$)}

The inference spend share $S_K$, one of the two components of $\varepsilon_{R,C}=\gamma S_K$.
\end{minipage} & \begin{minipage}[t]{\linewidth}\raggedright\setlength{\parskip}{1em}
As aforementioned in the last subsection, we have almost no data on lab expenditure on inference or its share relative to other R\&D inputs.
\end{minipage} & \begin{minipage}[t]{\linewidth}\raggedright\setlength{\parskip}{1em}
\textbf{Ideal new data}: Labs could publish detailed data on inference compute spend across domains and over time.
\end{minipage} \\
\addlinespace[1.2em]
\begin{minipage}[t]{\linewidth}\raggedright\setlength{\parskip}{1em}
\textbf{Tradeoff between training and inference compute ($\gamma$)}

The quality--quantity exponent $\gamma$, the other component of $\varepsilon_{R,C}=\gamma S_K$.
\end{minipage} & \begin{minipage}[t]{\linewidth}\raggedright\setlength{\parskip}{1em}
The parameter $\gamma$ governs the returns to scaling capabilities $C$ versus scaling inference compute $K$, since we model effective AI labor as $C^\gamma K$. Estimates from \citet{villalobos2023trading} imply $\gamma \approx 0.15$--$0.3$ under the ECI normalization of capabilities ($C=e^{\text{ECI}}$).\tablefootnote{\citet{villalobos2023trading} estimate the training--inference compute tradeoff, which in our notation is $\gamma \times \varepsilon_{C,T} \approx 1$ or $2$. Dividing by $\varepsilon_{C,T}\approx 6.5$ from Section~\ref{subsec:capabilities-to-algo} gives $\gamma$.}
\end{minipage} & \begin{minipage}[t]{\linewidth}\raggedright\setlength{\parskip}{1em}

\textbf{Current data weakness}: The estimation is not up to date, and the implied $\gamma$ divides estimates from two separate studies with different tasks and model generations.

\textbf{Ideal new data}: More extensive study on public benchmarks/with ECI.
\end{minipage} \\
\addlinespace[1.2em]
\begin{minipage}[t]{\linewidth}\raggedright\setlength{\parskip}{1em}
\textbf{The effect of AI use on researcher productivity}

A direct (if loose) measure of $\varepsilon_{R,C}$: how much capability raises effective R\&D.
\end{minipage} & \begin{minipage}[t]{\linewidth}\raggedright\setlength{\parskip}{1em}
We have self-reported productivity gains from surveys: 1.4--2X in METR's survey of technical workers \citep{becker2026aiusage}, and roughly 4X in the Claude Mythos Preview system card \citep{anthropic2026mythospreview}. The magnitudes are hard to interpret, but the upward trend is clear.

\citet{favaro2026rsi} report lines of code merged per engineer per day rose roughly 8X between 2024 and Q2 2026.
\end{minipage} & \begin{minipage}[t]{\linewidth}\raggedright\setlength{\parskip}{1em}
\textbf{Current data weakness}: Note that this is the productivity effect of AI, but $\varepsilon_{R,C}$ is the total research effort added due to one extra unit of capability.

\textbf{Ideal new data}: An RCT that measures productivity benefits from extra capabilities.
\end{minipage} \\
\addlinespace[1.2em]
\begin{minipage}[t]{\linewidth}\raggedright\setlength{\parskip}{1em}
\textbf{The strength of autonomous R\&D}

A direct measure of $\varepsilon_{R,C}$ (and the discovery-rate response $\varepsilon_{\dot A,C}$): AI's standalone contribution to discovery.
\end{minipage} & \begin{minipage}[t]{\linewidth}\raggedright\setlength{\parskip}{1em}
AI agents have recently advanced the frontier in some optimization problems; however, it is difficult to find a metric to measure their overall ability.\tablefootnote{See \citet{cunningham2026applepicking} for discussion of recent autonomous AI R\&D ability.}

Recent system cards report mixed results on benchmarks of AI R\&D. Models generally outperform humans on optimization problems when the humans are given between 8 and 40 hours, but show low pass rates on complex debugging problems (Claude Mythos Preview system card, \citealp[Table 2.3.3.A]{anthropic2026mythospreview}; GPT-5.5 system card, \href{https://deploymentsafety.openai.com/gpt-5-5/ai-self-improvement}{\uline{section 9.1.3}}). These results are difficult to place on a common scale.

\end{minipage} & \begin{minipage}[t]{\linewidth}\raggedright\setlength{\parskip}{1em}
\textbf{Current data weakness}: Benchmark questions are not the same as real-life LLM training problems.

\textbf{Ideal new data}: How autonomous AI systems perform on real LLM training problems would be especially informative.\\
The ideal data would characterize the returns to expenditure on agentic, human, and hybrid labor.
\end{minipage} \\
\addlinespace[1.2em]
\begin{minipage}[t]{\linewidth}\raggedright\setlength{\parskip}{1em}
\textbf{Capabilities of AI R\&D}

A direct, qualitative read on $\varepsilon_{R,C}$ --- which parts of research AI substitutes for, and where it remains weak.
\end{minipage} & \begin{minipage}[t]{\linewidth}\raggedright\setlength{\parskip}{1em}
\citet{si2025novel} find that AI-generated AI research ideas are rated highly relative to human-generated ones. \citet{trehan2026llms} find a range of weaknesses, including both taste and execution.

The Claude Mythos Preview system card has an extended discussion of the model's qualitative strengths and weaknesses in AI R\&D \citep{anthropic2026mythospreview}.
\end{minipage} & \begin{minipage}[t]{\linewidth}\raggedright\setlength{\parskip}{1em}
\textbf{Current data weakness}: It remains unclear how to describe AI's weaknesses in AI R\&D relative to human researchers. A common split is between idea generation (taste) and idea execution.

\textbf{Ideal new data}: Unstructured interviews with lab staff with well-designed questions eliciting AI R\&D weaknesses. 
\end{minipage} \\\\
\bottomrule
\end{longtable}

\begin{tcolorbox}[
breakable,
  colback=black!5,
  colframe=black!70,
  title=Technical details: research effort to capabilities,
  fontupper=\fontsize{12pt}{14pt}\selectfont,
  fonttitle=\bfseries
]
Suppose AI enters research effort through a capability-adjusted count of effective researchers, $N = K\,C^{\gamma}$, where $K$ is the quantity of inference compute and $C^{\gamma}$ is a quality adjustment, so that $R = R(\dots, N, \dots)$. By the chain rule,
\[
\varepsilon_{R,C} = \frac{\partial \log R}{\partial \log C}
= \frac{\partial \log R}{\partial \log N}\,\frac{\partial \log N}{\partial \log C}
= \varepsilon_{R,N}\cdot \gamma,
\]
since $\partial \log N/\partial \log C = \gamma$. Effective researchers are purchased through inference compute, so the expenditure on this input is the inference-compute spend and its cost share is $S_K$. Applying the cost-share correspondence (see the technical details on cost minimization above), $\varepsilon_{R,N}=S_K$, giving
\[
\varepsilon_{R,C} = \gamma\, S_K.
\]
\end{tcolorbox}

\subsection{Capabilities Response to Algorithmic Efficiency}
\label{subsec:capabilities-to-algo}

The third term is the elasticity of capabilities with respect to algorithmic progress. Suppose that algorithmic efficiency and training compute enter $C$ symmetrically (both through effective compute, $C = C(A\times T)$). This implies that the elasticity of capabilities to algorithmic progress equals the elasticity of capability to training compute i.e. $\varepsilon_{C,A}=\varepsilon_{C,T}$. Note that $\varepsilon_{C, T}$ is precisely the object estimated by scaling-law studies. Epoch's work on the capabilities--compute relationship \citep{ho2025rosetta, epoch2025eci} estimates values of $\varepsilon_{C, T}$ around $6.5$ when $C$ is defined as the exponential of the Epoch Capabilities Index, $e^{\text{ECI}}$.\footnote{The magnitude depends on this normalization of capabilities; the loop gain $\varepsilon_{R,C}\,\varepsilon_{C,A}$ that enters the condition is invariant to it.}

One qualification is needed: the symmetry $\varepsilon_{C,A}=\varepsilon_{C,T}$ relies on algorithmic progress being scale-free, so that it enters capability only through the product $A\times T$. \citet{gundlach2025origin} present evidence that algorithmic progress is instead scale-biased (faster at frontier scale than at small scale for pretraining), in which case $\varepsilon_{C,A}$ and $\varepsilon_{C,T}$ can diverge and scaling-law estimates identify only the latter. Nonetheless, we treat this term as the best-measured of the four; the supporting model-training data is collected in the supplementary material (Section~\ref{subsec:supplementary}).

\begin{longtable}[]{@{}
  >{\raggedright\arraybackslash}p{(\columnwidth - 4\tabcolsep) * \real{0.21}}
  >{\raggedright\arraybackslash}p{(\columnwidth - 4\tabcolsep) * \real{0.44}}
  >{\raggedright\arraybackslash}p{(\columnwidth - 4\tabcolsep) * \real{0.36}}@{}}
\toprule
\begin{minipage}[b]{\linewidth}\raggedright
\textbf{Model primitive}
\end{minipage} & \begin{minipage}[b]{\linewidth}\raggedright
\textbf{Existing data}
\end{minipage} & \begin{minipage}[b]{\linewidth}\raggedright
\textbf{Data ask}
\end{minipage} \\
\midrule
\endhead
\begin{minipage}[t]{\linewidth}\raggedright\setlength{\parskip}{1em}
\textbf{Capabilities' elasticity to algorithmic progress ($\varepsilon_{C,A}$)}
\end{minipage} & \begin{minipage}[t]{\linewidth}\raggedright\setlength{\parskip}{1em}
\citet{ho2025rosetta} and \citet{epoch2025eci} estimate the capability--compute relationship, implying $\varepsilon_{C,A}\approx 6.5$ when $C=e^{\text{ECI}}$.
\end{minipage} & \begin{minipage}[t]{\linewidth}\raggedright\setlength{\parskip}{1em}
\textbf{Current data weakness}: Estimates assume scale-free algorithmic progress; \citet{gundlach2025origin} suggest it is scale-biased.

\textbf{Ideal new data}: Algorithmic efficiency estimates at multiple scales, to separate $\varepsilon_{C,A}$ from $\varepsilon_{C,T}$.
\end{minipage} \\
\bottomrule
\end{longtable}

\begin{tcolorbox}[
breakable,
  colback=black!5,
  colframe=black!70,
  title=Technical details: Measuring the superelasticity term,
  fontupper=\fontsize{12pt}{14pt}\selectfont,
  fonttitle=\bfseries
]
\subsection{The Superelasticity Term}
\label{subsec:superelasticity}

As described in footnote \ref{fn:super}, the condition for a self-sustaining acceleration in capabilities \eqref{eq:measurement_condition} technically has a fourth term. The fourth term asks whether the capabilities elasticity $\varepsilon_{C,A}$ itself changes as algorithmic efficiency grows, i.e. whether $\mathrm d\log \varepsilon_{C,A}/\mathrm d\log A \neq 0$. Our best guess is that it is close to zero. As we showed in the previous subsection, $\varepsilon_{C,A}=\varepsilon_{C,T}$ under certain conditions. Thus the superelasticity measures the proportional change in $\varepsilon_{C,T}$ as effective compute grows, which depends on the curvature of the capability--compute relationship rather than its slope. That relationship has looked close to log-linear across many orders of magnitude of training compute \citep{ho2025rosetta, epoch2025eci}, so $\varepsilon_{C,T}$ appears approximately constant and therefore the superelasticity near zero.

\vspace{1em}

\noindent
The caveat is the same scale-free assumption as in the previous subsection: if algorithmic progress is scale-biased \citep{gundlach2025origin}, the symmetry breaks, and constancy of $\varepsilon_{C,T}$ need not imply constancy of $\varepsilon_{C,A}$.

\begin{longtable}[]{@{}
  >{\raggedright\arraybackslash}p{(\columnwidth - 4\tabcolsep) * \real{0.21}}
  >{\raggedright\arraybackslash}p{(\columnwidth - 4\tabcolsep) * \real{0.44}}
  >{\raggedright\arraybackslash}p{(\columnwidth - 4\tabcolsep) * \real{0.36}}@{}}
\toprule
\begin{minipage}[b]{\linewidth}\raggedright
\textbf{Model primitive}
\end{minipage} & \begin{minipage}[b]{\linewidth}\raggedright
\textbf{Existing data}
\end{minipage} & \begin{minipage}[b]{\linewidth}\raggedright
\textbf{Data ask}
\end{minipage} \\
\midrule
\endhead
\begin{minipage}[t]{\linewidth}\raggedright\setlength{\parskip}{1em}
\textbf{Superelasticity of capabilities ($\frac{\mathrm d\log \varepsilon_{C,A}}{\mathrm d\log A}$)}
\end{minipage} & \begin{minipage}[t]{\linewidth}\raggedright\setlength{\parskip}{1em}
No direct estimates, but the capability--compute relationship appears close to log-linear across many orders of magnitude of compute \citep{ho2025rosetta, epoch2025eci}, which suggests $\varepsilon_{C,T}$, and by symmetry $\varepsilon_{C,A}$, is approximately constant.
\end{minipage} & \begin{minipage}[t]{\linewidth}\raggedright\setlength{\parskip}{1em}
\textbf{Current data weakness}: Scale-biased algorithmic progress would break the symmetry with $\varepsilon_{C,T}$.\\
\textbf{Ideal new data}: Capability--compute estimates spanning a wide range of effective compute, updated past 2024, to estimate curvature and not just slope.
\end{minipage} \\
\bottomrule
\end{longtable}
\end{tcolorbox}

\subsection{Supplementary Data}\label{subsec:supplementary}

The data below do not measure the three terms in \eqref{eq:measurement_condition} directly but fill out the broader picture of AI R\&D automation, including economic feedback loops. For example, the model-training laws row helps pin down the third term, the capability--compute scaling relationship (Section~\ref{subsec:capabilities-to-algo}). The same data also help identify historical growth in algorithmic efficiency and predict the effects of future growth in training compute and data. The final row covers the economic feedback loop, through which capability funds further inputs.

\begin{longtable}[]{@{}
  >{\raggedright\arraybackslash}p{(\columnwidth - 4\tabcolsep) * \real{0.37}}
  >{\raggedright\arraybackslash}p{(\columnwidth - 4\tabcolsep) * \real{0.32}}
  >{\raggedright\arraybackslash}p{(\columnwidth - 4\tabcolsep) * \real{0.31}}@{}}
\toprule
\begin{minipage}[b]{\linewidth}\raggedright
\textbf{Model primitive}
\end{minipage} & \begin{minipage}[b]{\linewidth}\raggedright
\textbf{Existing data}
\end{minipage} & \begin{minipage}[b]{\linewidth}\raggedright
\textbf{Data ask}
\end{minipage} \\
\midrule
\endhead
\begin{minipage}[t]{\linewidth}\raggedright\setlength{\parskip}{1em}
\textbf{Model training laws}

Calibrates the capability map $C(AT)$; pins down $\varepsilon_{C,A}=\varepsilon_{C,T}$.
\end{minipage} & \begin{minipage}[t]{\linewidth}\raggedright\setlength{\parskip}{1em}
The basic model remains Chinchilla \citep{hoffmann2022training}, which maps data and compute to pretraining loss.
\end{minipage} & \begin{minipage}[t]{\linewidth}\raggedright\setlength{\parskip}{1em}
\textbf{Current data weakness}: Scaling laws for RL remain largely unpublished.

\textbf{Ideal new data}: post-training compute usage of models. 
\end{minipage} \\
\addlinespace[1.2em]
\begin{minipage}[t]{\linewidth}\raggedright\setlength{\parskip}{1em}
\textbf{Growth in training compute}

$g_T$; since effective compute is $AT$, it drives capability growth.
\end{minipage} & \begin{minipage}[t]{\linewidth}\raggedright\setlength{\parskip}{1em}
Training compute has grown 4--5X per year since 2010 \citep{rahman2024compute}. \citet{whitfill2025forecasting} discusses possible GDP constraints on continued compute scaling.
\end{minipage} & \begin{minipage}[t]{\linewidth}\raggedright\setlength{\parskip}{1em}
\textbf{Current data weakness}: There is scarce data on model training compute since 2024.

\textbf{Ideal new data}: training compute of new models. 
\end{minipage} \\
\addlinespace[1.2em]
\begin{minipage}[t]{\linewidth}\raggedright\setlength{\parskip}{1em}
\textbf{Growth in data}

Data $D$, an input to capability beyond effective compute $AT$.
\end{minipage} & \begin{minipage}[t]{\linewidth}\raggedright\setlength{\parskip}{1em}
There are minimal existing estimates of data spend by labs.\tablefootnote{This need not preclude an explosion in capabilities: broad-based data collection (e.g., via workplace surveillance) might interact with deployment of AI systems across the economy to generate a `data feedback loop'. However, we expect this to only occur at longer timescales.}
\end{minipage} & \begin{minipage}[t]{\linewidth}\raggedright\setlength{\parskip}{1em}
\textbf{Ideal new data}: Labs could publish data spend, or partial estimates could be constructed from third-party data providers such as Surge AI or Scale AI. 
\end{minipage} \\
\addlinespace[1.2em]
\begin{minipage}[t]{\linewidth}\raggedright\setlength{\parskip}{1em}
\textbf{Share of AI's economic value creation captured by AI labs}

Economic output $Y$, the channel through which capability funds further inputs (the economic feedback loop).
\end{minipage} & \begin{minipage}[t]{\linewidth}\raggedright\setlength{\parskip}{1em}
We can use firm revenue estimates, but these are likely to represent only a small share of the value created. \citet{brynjolfsson2026worth} use online choice experiments to elicit willingness-to-accept for giving up generative AI chatbots, estimating that consumer surplus substantially exceeds firms' revenues.
\end{minipage} & \begin{minipage}[t]{\linewidth}\raggedright\setlength{\parskip}{1em}
\textbf{Current data weakness}: More estimation needed, with time variation.

\textbf{Ideal new data}: AI firms' revenue over time.  
\end{minipage} \\\\

\begin{minipage}[t]{\linewidth}\raggedright\setlength{\parskip}{1em}
\textbf{Elasticity of GDP to capabilities ($\varepsilon_{Y,C}$)}

How much a given capability improvement raises aggregate output.
\end{minipage} & \begin{minipage}[t]{\linewidth}\raggedright\setlength{\parskip}{1em}
Little direct evidence: AI's measured contribution to GDP remains small relative to its measured capabilities. \citet{yotzov2026firm} survey over 5000 executives: $9/10$ report no impact of AI on their firm's productivity or employment over the past three years, but on average they expect AI to raise firm productivity by 1.4\% and output by 0.8\% over the next three years.
\end{minipage} & \begin{minipage}[t]{\linewidth}\raggedright\setlength{\parskip}{1em}
\textbf{Current data weakness}: Adoption and usage data are not tightly linked to impact on output.

\textbf{Ideal new data}: Task-level adoption and usage data linked to firm or sector output, tracking how output effects grow as capabilities improve.
\end{minipage}  \\\\

\begin{minipage}[t]{\linewidth}\raggedright\setlength{\parskip}{1em}
\textbf{Elasticity of AI firm budget to capabilities ($\varepsilon_{Y,C}$)} \\ 
We might instead interpret $Y$ as the budget AI firms have to spend on data, compute, and researchers. Then, $\varepsilon_{Y,C}$ is how much additional budget a given capability improvement generates (via fundraising or revenues).
\end{minipage} & \begin{minipage}[t]{\linewidth}\raggedright\setlength{\parskip}{1em}
Frontier-lab revenues have grown around 3x/year \citep{epoch2026aicompanies} with notable outliers, e.g. Anthropic. But lab expenditure is largely financed by external investment rather than revenue so the loop runs through expectations. 

\end{minipage} & \begin{minipage}[t]{\linewidth}\raggedright\setlength{\parskip}{1em}
\textbf{Current data weakness}: Revenue is observed but at low frequency. Lab budgets are largely financed by investor expectations rather than current revenue. 

\textbf{Ideal new data}: Credible estimates for how lab financing costs (through equity or debt) respond to changes in capabilities progress.
\end{minipage} \\\\
\bottomrule
\end{longtable}
\ifSubfilesClassLoaded{%
    \paperbibliography%
}{}

\end{document}

\section{Discussion}
 
To assess the plausibility of self-sustaining acceleration, we provide a very rough calibration of our analytical condition with the available estimates from Section \ref{sec:data}. Certain parameters are unknown, especially the elasticity of R\&D to AI capabilities. Thus, we derive a condition on the range of values for this elasticity that implies self-sustaining acceleration. We have a high degree of uncertainty in many of these empirical estimates, and also acknowledge the possibility of model misspecification. Thus we additionally provide a general discussion of the most important points of evidence for and against self-sustaining acceleration.

\subsection{Model calibration} 
\label{sec:modelcalibration}

We now use the estimates we have from Section \ref{sec:data} to calibrate the self-sustaining acceleration condition \eqref{eq:measurement_condition}:

\[
\underbrace{
\frac{\varepsilon_{\adot A,R}}{1-\varepsilon_{\adot A,A}}
}_{\text{return to R\&D}}
\cdot
\underbrace{
\varepsilon_{R,C}
}_{\substack{\text{research effort's} \\ \text{elasticity to} \\ \text{AI capabilities}}}
\cdot
\underbrace{
\varepsilon_{C,A}
}_{\substack{\text{capabilities'} \\ \text{elasticity to} \\ \text{algorithmic} \\ \text{efficiency}}}
>
1.
\]

We plug in the following values:
\begin{itemize}
    \item $\frac{\varepsilon_{\dot{A},R}}{1-\varepsilon_{\dot{A},A}} = \frac{g_A}{g_R} \approx \frac{\ln(3)}{\ln(3)} = 1$. We take $g_A = \ln(3)$ as the annual growth rate in algorithmic progress per \citet{ho2024algorithmic}, meaning $A$ triples each year. Next recall $g_R = S_L g_L + S_E g_E + S_K g_K$. We take $g_L \approx \ln 3$ from \citet{epoch2026aicompanies, whitfill2025compute} and $g_E \approx \ln(3)$ from \citet{epoch2026aichipproduction}. $g_K$ was probably small until the rise of coding agents but has since increased. Indeed OpenAI reported in the GPT-5.6 launch post in July 2026 that ``over the past six months, the share of research compute devoted to internal coding inference grew 100-fold,'' implying a much larger $g_R$ at present. But since we do not know the spend shares on these factors, and $S_K$ is likely still low relative to $S_L$ and $S_E$, we assume $g_R = \ln(3)$ for now. This assumption is further justified by the fact that $g_A$ was estimated in 2024. 
    \item $\varepsilon_{R,C}$. We have almost no evidence for this parameter because AI uplift studies do not report capability improvements in a form that maps cleanly to this elasticity.
    \item $\varepsilon_{C,A} \approx 6.5$. Define capabilities as the exponential of the Epoch Capabilities Index, $\exp(\text{ECI})$. Also recall the assumption that $C=C(A \times T)$ which implies $\varepsilon_{C,A} = \varepsilon_{C,T}$. Thus we can get the elasticity of capabilities to algorithmic efficiency from the slope of a regression of ECI on $\log(\text{compute})$ across models using data from Epoch AI which yields an estimate of roughly 6.5.\footnote{Setting aside the scale-bias in algorithmic progress, discussed below.}
\end{itemize}

Plugging in these values tells us that self-sustaining acceleration is achieved if $\varepsilon_{R,C} > 0.15$. This is the increase in R\&D effort per additional unit of ECI. For comparison, Claude Opus 4.8 scores one ECI unit more than Claude Opus 4.7. Thus under our model, Anthropic would meet the self-sustaining acceleration condition if adopting Claude Opus 4.8 led to 15\% higher research productivity than Opus 4.7. Note that the increase in research output per unit of AI capabilities tends to be smaller than the uplift for human tasks from AI use, because labor only accounts for a fraction of all research tasks. 

It is hard to say whether one unit of the ECI is worth 15\% more research effort, but a rough calculation based on the last two years of AI progress reassuringly suggests that the returns to capabilities are currently lower than this threshold. Claude Opus 4.8 scores 16 ECI points more than Claude 3.7 Sonnet, which was released in February 2025 alongside Claude Code. At the time, estimates for AI uplift on engineering tasks varied but were not conclusively larger than 0.\footnote{For example, an influential study from \citet{metr-2025-early-2025-ai-experienced-os-dev-study} found a negative impact of AI on engineering productivity.} More recently, Anthropic engineers surveyed in the Claude Mythos Preview system card reported a 4X uplift from Claude use \citep{anthropic2026mythospreview}. The 4X uplift is very likely to be an overestimate, but even if it were true it would imply a 9\% increase in productivity per unit of ECI, below the threshold for self-sustaining acceleration over this period.

Of course, these elasticities are likely increasing -- the move from GPT-4 to GPT-5 led to a much larger increase in research productivity than the move from GPT-1 to GPT-2. The continual measurement of those elasticities is necessary for understanding the current state of RSI. Understanding how and why $\varepsilon_{R,C}$ changes over time is therefore an important research priority.\footnote{One promising direction here is to model automation as endogenous: as AI capabilities improve, which research tasks do labs choose to automate, and how do those choices affect the aggregate responsiveness of research effort to capabilities?}

\subsection{Arguments for and against acceleration} 

Given the uncertainty in the above calibration, we now turn to general qualitative arguments for and against acceleration, informed by our models and data.

\paragraph{Best evidence in favor of acceleration:}
\begin{itemize}

\item AI systems are meaningfully contributing to AI progress. The Claude Mythos Preview system card reported a self-assessed productivity uplift of roughly 4X among Anthropic researchers (although there are reasons to think this is likely overstated), and a 40-hour time horizon at which Claude Mythos Preview beat human researchers \citep{anthropic2026mythospreview}. \citet{favaro2026rsi} assess the performance of human researchers versus Claude Code in making AI research decisions. In April 2026, Mythos Preview beat humans 64\% of the time, up from 50\% for Claude models released in 2025. OpenAI reported in July 2026 that internal coding inference's share of research compute grew 100-fold in the prior six months, suggesting that the value of AI in R\&D has dramatically increased. 



\item In August 2025, experts and superforecasters \citep{metr2025forecasting}\footnote{``Experts and superforecasters both find a three-fold acceleration of effective compute scale-up by 2029 plausible but unlikely, with experts giving it a median 20\% chance and superforecasters 8\%.''} predicted an 8--20\% chance that the growth rate of effective compute would triple by 2029. In May 2026, Jack Clark predicted a 60\% chance that by the end of 2028, there will be ``an AI system powerful enough that it could autonomously build its own successor'' \citep{clark2026importai}.

\end{itemize}

\paragraph{Best evidence against acceleration:}
\begin{itemize}
\item Compute bottlenecks could bind for multiple reasons. Algorithmic improvements may require increases in training compute \citep{gundlach2025origin}. Cognitive labor increases may require increases in experimental compute \citep{whitfill2025compute}. Productivity growth may not be fast enough to support continued AI R\&D investments, that is, we ``run out of GDP'' \citep{halperin2026}. 
\item Autonomous contributions to algorithm optimization may quickly hit diminishing returns. This has been widely observed informally and formalized in the apple-picking model of AI R\&D developed by \citet{cunningham2026applepicking}.
\item Capabilities may depend on data. \citet{millidge2025data} argues that ``most algorithmic progress is data progress''. This includes everything from better filtering of pre-training data to the substantial resources spent on post-training data. Pre-training filtering seems to be a one-time boost, whereas post-training data might be supplied more elastically and be amenable to automation. We are only beginning to learn how to build better reinforcement learning environments, and AI systems could help design future ones. 
\end{itemize}

\paragraph{Narrow capabilities acceleration before broad capabilities acceleration.}
It seems plausible that AI R\&D agents, while effective at making concrete algorithmic progress as captured by benchmarks, might remain less effective at developing major breakthroughs that enable models to learn efficiently in new environments, or develop more amorphous capabilities like `vision' or `taste'. This idea would be captured by combining our models in Sections \ref{sec:narrow-broad-capabilities} and \ref{sec:spurts} so that the feedback loop from narrow capabilities is helpful for generating narrow algorithmic progress but not broad progress. Nonetheless, spillovers from narrow to broad progress remain possible. For example, an advanced narrow system, in pursuit of benchmark optimization, may discover new capabilities, such as highly sample-efficient algorithms able to update their weights during deployment, that quickly turn narrow progress into broader progress.

\paragraph{Political and social barriers.}
Even if a self-sustaining acceleration is technically plausible, political and social barriers could stop it in its tracks. In America, at least 20 data center projects were canceled in the first three months of 2026 due to local pushback; \$85 billion worth of projects have been canceled in the last three years \citep{economist2026datacentrebacklash}. Recent policy reactions to Mythos by the Trump administration, despite an earlier laissez-faire stance, also suggest placing greater weight on these barriers. If compute scaling remains important, even minor frictions in model deployment will slow revenue growth and further scaling.

\subsection{Impacts of a capabilities acceleration on the world}

A self-sustaining acceleration in AI progress differs from generic AI progress in \emph{speed}: it could compress generations of progress into short periods of time. Indeed, if self-sustaining progress were feasible, it is likely that the returns to increasing investment in inputs like compute and data would be high, further increasing investment and the pace of progress. We highlight several aspects of rapid progress that distinguish it from gradual AI progress, and that may be especially important for understanding its social, economic, and political consequences.

\begin{itemize}
\item \textbf{Transition costs:} It seems plausible that transition costs increase with the speed of technological change. Labor markets are one such case: the costs of structural transformation may rise with its speed, as some economists argue was the case with the China Shock.
\item \textbf{Institutional inertia:} In a world of rapidly accelerating capabilities growth, institutions broadly defined --- regulatory capacity, democratic procedures, legal frameworks, infrastructure, and so on --- might lag behind \citep{bostrom2014superintelligence, ord2020precipice}. Rapid technological change might also grant political elites a temporary spike in \emph{de facto} power, which they could use to capture institutions. 
\item \textbf{Asymmetric progress across domains:} As noted, we see self-sustaining acceleration in narrow capabilities as more likely in the near term than an acceleration in broad capabilities. The former would likely be concentrated in tasks that are easy to verify in closed loops, such as software development, games, mathematics, and related areas. This could create acute asymmetries in progress across domains. For example, we may find solutions to long-standing puzzles in mathematics and breakthroughs in computational chemistry that enable new materials, alongside sluggish progress in less self-contained or messier domains.
\item \textbf{Power imbalances across firms and nation-states:} Even model providers with relatively small gaps in technical skill could see large gaps open up in effective capabilities and power over the world. Imagine Anthropic or the US government having access to capabilities that are vastly superior to those of any other actor.
\item \textbf{Greater misalignment risk:} Misaligned AI might also create bigger risks under a capabilities acceleration. Sudden increases in model capabilities would provide less time for technical and social safeguards to be put in place.
\end{itemize}

\subsection{Future Steps}
There are several directions for future work that build off this note, some of which we are actively pursuing:
\begin{itemize}
    \item Empirically estimating the self-sustaining acceleration condition for the model with economic feedback loops.
    \item Measuring the return to research off the balanced growth path.
    \item Combining models with multiple types of AI algorithms and capabilities (Sections~\ref{sec:narrow-broad-capabilities} and~\ref{sec:spurts}).
    \item Distinguishing quality from quantity improvements in data and compute.
    \item Extending the model to multiple research sectors (software, hardware, data).
    \item Modeling research taste and vision in AI R\&D.
    \item Endogenizing the decision to automate AI R\&D.
    \item Incorporating directed technical change.
    \item Surveys and field experiments within AI frontier labs to improve estimates of the key parameters in the model.
    \item Developing new ways to incentivize participation in prediction markets to produce better forecasts of AI progress (\cite{fradkinjabariankoh}).
\end{itemize}

\ifSubfilesClassLoaded{%
    \paperbibliography%
}{}

\end{document}

\paperbibliography

\appendix
\newpage
\section*{\LARGE Appendix}

\section{Scale-Dependent Algorithmic Progress} \label{app:scaledep}

Since algorithmic progress is usually not directly observable, it is typically estimated as the increase in capabilities not explained by increases in training compute, just as TFP is estimated by the Solow residual. That is, if we know (from a ``scaling law'') that in the absence of algorithmic progress some index of AI capabilities $C$ would track training compute $T$, it is conventional to propose the model
\begin{align}
    C = A \cdot T \label{zequalsat}
\end{align}
and calculate $A$ as $C/T$. Using a model like our ``baseline model'' in Section \ref{sec:simplemodel}, the resulting time series for $A$, in combination with a time series for research inputs $R$ and the semi-endogenous AI R\&D function, can then be used to generate a prediction for how fast algorithms would advance after the automation of AI R\&D.

\citet{gundlach2025origin} show conclusively that model (\ref{zequalsat}) is misspecified because algorithmic progress in AI development to date has been largely biased toward large scales. For an extreme case of large-scale-biased algorithmic progress, consider instead the model
\begin{align}
    C = \min(A,T)\cdot T. \label{zequalsminatt}
\end{align}
Here, algorithmic improvements proportionally advance AI capabilities as long as there is the training compute to leverage them (i.e., here, as long as $T>A$). If we stopped scaling the compute, however, capabilities would ultimately stagnate (here, at $T^2$). Likewise, even if automating AI R\&D yields a leap in the quality of the algorithms available, under (\ref{zequalsminatt}) this will not deliver arbitrarily large advances in AI capabilities until we have accumulated the training compute to leverage these algorithms.

Another potential explanation for \citeauthor{gundlach2025origin}'s finding, however, is that we have been developing large-scale-biased algorithms precisely \textit{because} we have been scaling compute. Indeed, we might imagine living in a somewhat different world in which we knew exactly what architectures would take us to superintelligence but were simply too compute-constrained to make efficient use of them. Why don't we live in that world? One answer is that we have no incentive to discover such algorithms---and would not be able to validate them if we did---in a world in which we are compute-constrained. This also suggests that if we faced a long-term compute constraint, or if an AI model were tasked with recursively self-improving on fixed compute, we might develop algorithms that made ever better use of ``small'' training runs---as academic labs do to some extent today. That is, rapid RSI may still be feasible, but only via some kind of ``directed technical change''.

\ifSubfilesClassLoaded{%
    \paperbibliography%
}{}

\end{document}

\section{Inference Compute} \label{app:inference-compute}

In the bottlenecks model (Section~\ref{sec:bottlenecks}), inference compute ($K$) enters algorithmic improvements directly. Here we instead treat inference scaling as its own sector of algorithmic progress. We relabel the algorithmic efficiency of the main text ($A$) as \textit{training efficiency} ($A_T$), since it combines with training compute ($T$) to produce capabilities ($C$). A second stock, \textit{inference efficiency} ($A_I$), combines with raw inference compute to yield \textit{effective inference compute} ($M = A_I K$). The two sectors sit in parallel, and both capabilities and effective inference compute flow into both kinds of improvements ($\adot{A}_T$ and $\adot{A}_I$). To keep the diagram simple we suppress experimental compute ($E$) and data ($D$), which enter as in the bottlenecks model.

\begin{figure}[h]
\captionof{figure}{Inference compute as a parallel sector of algorithmic progress}\label{fig:inference-compute-sector}
\centering
\scalebox{\rsiscale}{%
\begin{tikzpicture}[rsi diagram, >={Stealth[]}]
    \rsiNode[align=center, thick, at={(3,-0.6)}]{research}{human R\&D\\labor ($L$)}
    \rsiNode[align=center, thick, minimum width=3.7cm, text width=3.5cm, at={(0,-4.2)}]{algorithms}{training\\efficiency ($A_T$)}
    \rsiNode[align=center, thick, minimum width=3.7cm, text width=3.5cm, anchor=south, at={([yshift=-0.8pt]algorithms.north)}]{adot}{training\\improvements ($\dot A_T$)}
    \rsiNode[align=center, thick, at={(0,-7.0)}]{capabilities}{capabilities\\($C$)}
    \rsiNode[draw=black, thick, align=center, at={(-4.3,-7.0)}]{compute}{training\\compute ($T$)}
    \rsiNode[draw=newblue, text=newblue, very thick, align=center, minimum width=3.7cm, text width=3.5cm, at={(6,-4.2)}]{algorithmsI}{inference\\efficiency ($A_I$)}
    \rsiNode[draw=newblue, text=newblue, very thick, align=center, minimum width=3.7cm, text width=3.5cm, anchor=south, at={([yshift=-0.8pt]algorithmsI.north)}]{adotI}{inference\\improvements ($\dot A_I$)}
    \rsiNode[draw=newblue, text=newblue, very thick, align=center, at={(6,-7.0)}]{effinference}{effective inference\\compute ($M$)}
    \rsiNode[draw=black, thick, align=center, at={(10.3,-7.0)}]{inference}{inference\\compute ($K$)}
    \draw[black, very thick, cg edge] (algorithms.south) -- (capabilities.north);
    \draw[newblue, very thick, cg edge] (algorithmsI.south) -- (effinference.north);
    \draw[black, very thick, cg edge] (compute.east) -- (capabilities.west);
    \draw[black, very thick, cg edge] (inference.west) -- (effinference.east);
    \draw[black, very thick, cg edge] (research.west) -| ($(adot.north)+(0.5,0)$);
    \draw[black, very thick, cg edge] (research.east) -| ($(adotI.north)+(-0.5,0)$);
    \draw[black, very thick, cg edge] (algorithms.west) -- (-2.7,-4.2) |- (adot.175);
    \draw[newblue, very thick, cg edge] (algorithmsI.east) -- (8.7,-4.2) |- (adotI.5);
    \draw[black, very thick, cg edge] ([yshift=5pt]capabilities.east) -- (2.3,-6.82) |- ([yshift=-7pt]adot.east);
    \draw[newblue, very thick, cg edge] ([yshift=5pt]effinference.west) -- (3.6,-6.82) |- ([yshift=-8pt]adotI.west);
    \draw[black, very thick, cg edge] ([yshift=-5pt]capabilities.east) -- (3.1,-7.18) |- ([yshift=8pt]adotI.west);
    \draw[newblue, very thick, cg edge] ([yshift=-9pt]effinference.west) -- (2.7,-7.32) |- ([yshift=7pt]adot.east);
    \node[draw, thick, circle, fill=white, inner sep=1pt] at (algorithms.north) {\scriptsize${\blacktriangledown}$};
    \node[draw=newblue, thick, circle, fill=white, inner sep=1pt] at (algorithmsI.north) {\scriptsize\textcolor{newblue}{${\blacktriangledown}$}};
\end{tikzpicture}
}
\end{figure}
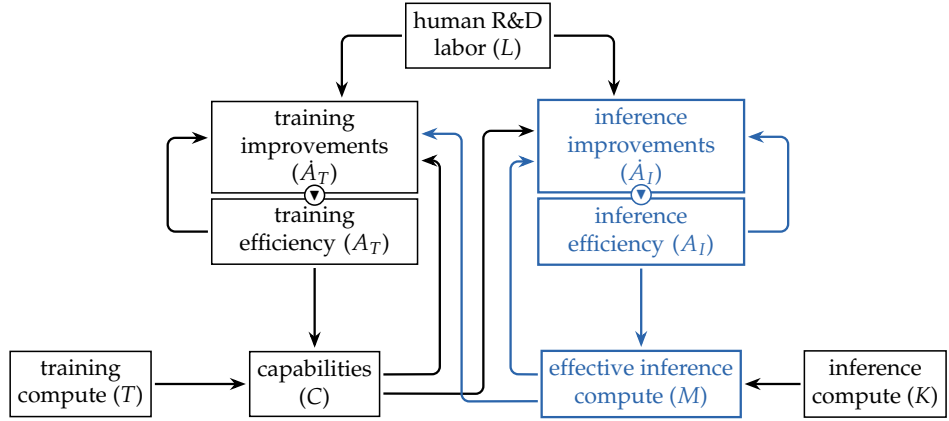

Formally, with $C = C(A_T,T)$ for capabilities, the two kinds of algorithmic improvements are produced as
\begin{align*}
    \adot{A}_T &= F(L,\, C,\, M;\, A_T), \\
    \adot{A}_I &= G(L,\, C,\, M;\, A_I).
\end{align*}

This adds two feedback loops to the baseline model. Inference efficiency has its own self-feedback loop ($A_I \to M \to \adot{A}_I$). It also creates a cross-sector loop: capabilities improve inference algorithms, and the resulting effective inference compute feeds back into training-efficiency research ($A_T \to C \to \adot{A}_I \to A_I \to M \to \adot{A}_T$).

\paragraph{Self-sustaining acceleration.} The condition for a self-sustaining acceleration in capabilities is that the sum of the two feedback channels exceeds one:
\begin{align}
\underbrace{
    \frac{
        \varepsilon_{\smash{\adot{A}_T},C}\,\varepsilon_{C,A_T}
    }{
        1-\varepsilon_{\smash{\adot{A}_T},A_T}
    }
}_{\text{core feedback loop}}
+
\underbrace{
    \frac{
        \varepsilon_{\smash{\adot{A}_T},M}\,\varepsilon_{M,A_I}\,
        \varepsilon_{\smash{\adot{A}_I},C}\,\varepsilon_{C,A_T}
    }{
        \left(1-\varepsilon_{\smash{\adot{A}_T},A_T}\right)
        \left(1-\varepsilon_{\smash{\adot{A}_I},A_I}-\varepsilon_{\smash{\adot{A}_I},M}\,\varepsilon_{M,A_I}\right)
    }
}_{\text{inference feedback loop}}
\; > \; 1.
\label{ssa:inference}
\end{align}
The first term is the core feedback loop of the baseline model. The second term is the new loop running through the inference sector. The condition reduces to the scalar condition \eqref{core_feedback_loop_condition} when the inference sector is shut down ($\varepsilon_{\smash{\adot{A}_T},M} = 0$). Because the second term enters additively, the inference loop can tip the system into a self-sustaining acceleration even when the core feedback loop alone falls short.

However, METR's expenditure horizon results suggest the inference loop is currently weak. \citet{cunningham2026expenditure} measure the returns to spending more inference compute on agents performing a frontier optimization task (speeding up NanoGPT training). Agents find optimizations cheaply at first, but their returns diminish much faster than the returns to expenditure on human labor: performance roughly plateaus in the low thousands of dollars per run. This is consistent with the apple-picking model of AI R\&D \citep{cunningham2026applepicking}, in which agents quickly exhaust the easy-to-find optimizations and returns to further search collapse. Both support the claim that $\varepsilon_{\smash{\adot{A}_T},M}$ falls quickly as inference expenditure scales up. If it falls to zero, the inference feedback loop vanishes and only the core feedback loop remains.
\begin{figure}
    \centering
    \includegraphics[width=\linewidth]{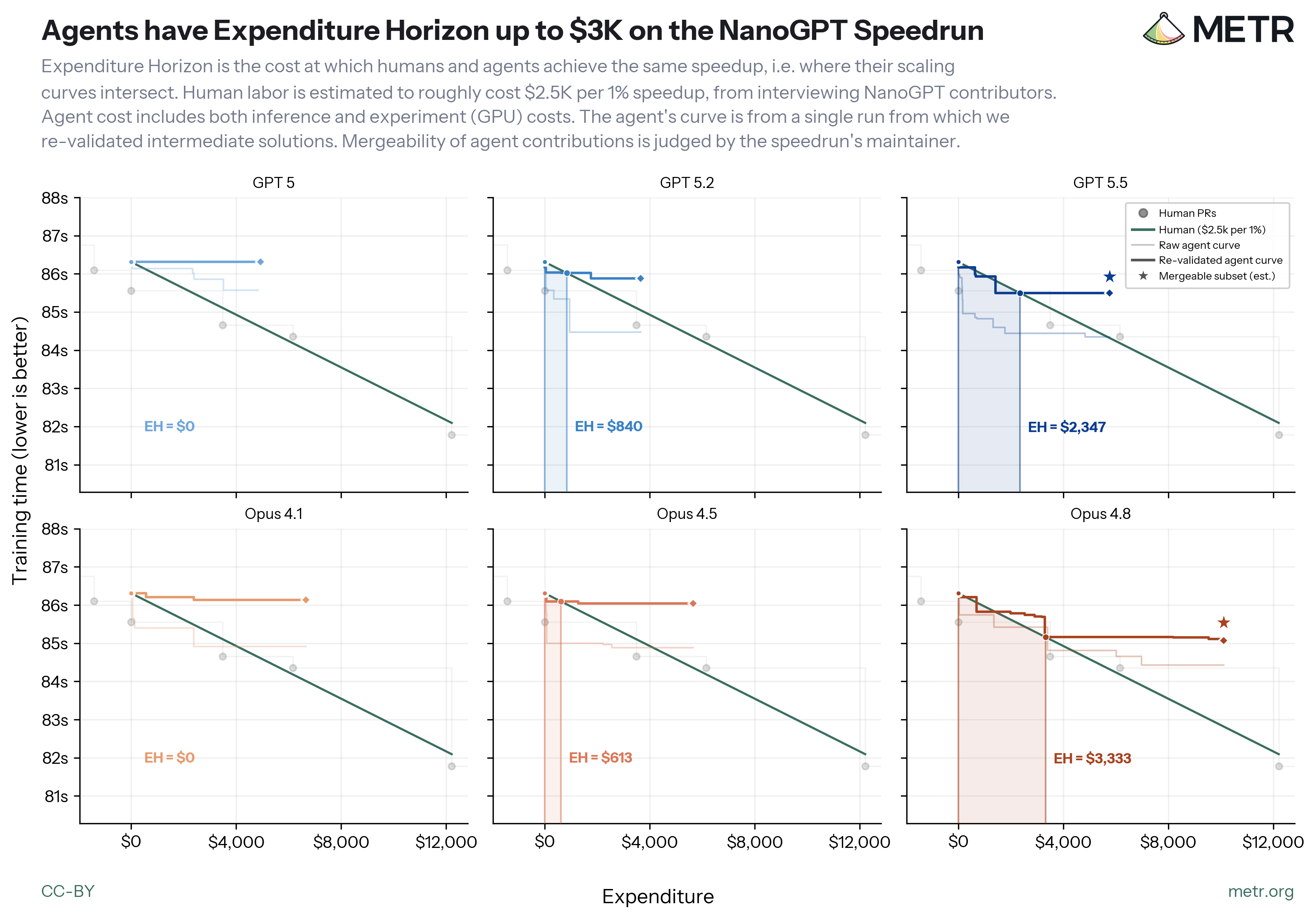}
    \label{fig:exp_horizon}
    \caption{\citet{cunningham2026expenditure}}
\end{figure}
\begin{tcolorbox}[
breakable,
  colback=black!5,      
  colframe=black!70, 
  title=Technical Details: Self-sustaining acceleration with two sectors,
  fontupper=\fontsize{12pt}{14pt}\selectfont,
  fonttitle=\bfseries
]
There are now two accumulating states, $X \equiv (A_T, A_I)'$. Holding $L$, $T$, and $K$ fixed, the reduced-form elasticity matrix $\bm{\mathcal{E}}$, with entries $\mathcal{E}_{ij} \equiv \frac{\mathrm d \log \dot{X}_i}{\mathrm d \log X_j}$, is
\begin{align}
\bm{\mathcal{E}} =
\begin{pmatrix}
    \varepsilon_{\smash{\adot{A}_T},A_T} + \varepsilon_{\smash{\adot{A}_T},C}\,\varepsilon_{C,A_T}
    & \varepsilon_{\smash{\adot{A}_T},M}\,\varepsilon_{M,A_I} \\[0.4em]
    \varepsilon_{\smash{\adot{A}_I},C}\,\varepsilon_{C,A_T}
    & \varepsilon_{\smash{\adot{A}_I},A_I} + \varepsilon_{\smash{\adot{A}_I},M}\,\varepsilon_{M,A_I}
\end{pmatrix}
\label{inference_loop_matrix}
\end{align}
The diagonal entries are each sector's own feedback loops; the off-diagonal entries are the cross-loops. As in the economic feedback loops of Section~\ref{sec:economicfeedback}, there is a self-sustaining acceleration when the dominant eigenvalue of $\bm{\mathcal{E}}$ exceeds one \citep{davidson2026automating}.

\vspace{1em}

For this $2\times 2$ system the eigenvalue condition has a closed form. Assuming neither sector's own loops are self-sustaining on their own ($\varepsilon_{\smash{\adot{A}_T},A_T} < 1$ and $\varepsilon_{\smash{\adot{A}_I},A_I} + \varepsilon_{\smash{\adot{A}_I},M}\,\varepsilon_{M,A_I} < 1$), the dominant eigenvalue of $\bm{\mathcal{E}}$ exceeds one exactly when $\det(\bm{I} - \bm{\mathcal{E}}) < 0$, i.e.\ when
\begin{align}
\left(1
- \varepsilon_{\smash{\adot{A}_T},A_T}
- \varepsilon_{\smash{\adot{A}_T},C}\,\varepsilon_{C,A_T}
\right)
\left(1 - \varepsilon_{\smash{\adot{A}_I},A_I} - \varepsilon_{\smash{\adot{A}_I},M}\,\varepsilon_{M,A_I}\right)
\; < \;
\varepsilon_{\smash{\adot{A}_T},M}\,\varepsilon_{M,A_I}\,
\varepsilon_{\smash{\adot{A}_I},C}\,\varepsilon_{C,A_T}.
\label{ssa:inference-det}
\end{align}
The left side multiplies the two sectors' `remaining gaps' after their own feedback loops; the right side is the product of elasticities around the cross-sector loop. Dividing both sides by the two gaps and moving the core loop across yields condition \eqref{ssa:inference}. Note that if $M = A_I K$ exactly, then $\varepsilon_{M,A_I}=1$ and the expressions simplify accordingly.

\vspace{1em}

The eigenvalue condition governs acceleration in the \textit{stocks} $(A_T, A_I)$, and is local: it takes the elasticities as given. While the cross-loop elasticities remain strictly positive, an acceleration in one stock pulls the other along. But if an elasticity falls over time (say $\varepsilon_{\smash{\adot{A}_T},M} \to 0$, so that better inference algorithms stop helping training research), the sectors decouple, and $A_I$ can accelerate without $A_T$.

\vspace{1em}

Since $C = C(A_T, T)$ and $T$ is held fixed, the growth rate of capabilities is proportional to the growth rate of training efficiency: $g_C = \varepsilon_{C,A_T}\, g_{A_T}$. With constant elasticities, capabilities accelerate exactly when training efficiency accelerates, so \eqref{ssa:inference} is also the condition for a self-sustaining acceleration in $C$. If the passthrough is not constant then, as in Section~\ref{sec:narrow-broad-capabilities}, the condition for acceleration in $C$ gains a superelasticity term $\frac{\mathrm d\log \varepsilon_{C,A_T}}{\mathrm d\log A_T}$ on the left side of \eqref{ssa:inference}. A negative superelasticity (for instance, a ceiling on model capabilities) can break the passthrough from accelerating algorithmic progress to accelerating capabilities.

\end{tcolorbox}

\ifSubfilesClassLoaded{%
    \paperbibliography%
}{}

\end{document}


\end{document}